\documentclass[]{rQFL2e}
\usepackage{booktabs,threeparttable}
\usepackage{color,multirow,multicol}

\bibpunct{(}{)}{,}{a}{}{,}
\begin{document}

\jvol{00} \jnum{00} \jyear{2013} \jmonth{January}

\markboth{Z.-Q. Jiang et al.}{Trading networks, abnormal motifs and stock manipulation}

\title{Trading networks, abnormal motifs and stock manipulation}

\author{
 Zhi-Qiang Jiang$\dag\ddag$,
 Wen-Jie Xie$\dag\ddag\S$,
 Xiong Xiong$\P$,
 Wei Zhang$^{\ast}\P$\thanks{$^\ast$Corresponding author. Email: weiz@tju.edu.cn},
 Yong-Jie Zhang$\P$,
 and
 Wei-Xing Zhou$^{\ast\ast}\dag\ddag\S$\thanks{$^{\ast\ast}$Corresponding author.  Email: wxzhou@ecust.edu.cn\vspace{12pt}}
 \\
\vspace{12pt}
\normalfont{
$\dag$School of Business, East China University of Science and Technology, Shanghai 200237, China\\
$\ddag$Research Center for Econophysics, East China University of Science and Technology, Shanghai 200237, China \\
$\S$Department of Mathematics, East China University of Science and Technology, Shanghai 200237, China\\
$\P$College of Management and Economics, Tianjin University, Tianjin 300072, China
}\\
\vspace{12pt}
\received{Submitted manuscript}
}

\maketitle

\begin{abstract}
  We study trade-based manipulation of stock prices from the perspective of complex trading networks constructed by using detailed information of trades. A stock trading network consists of nodes and directed links, where every trader is a node and a link is formed from one trader to the other if the former sells shares to the latter. Specifically, three abnormal network motifs are investigated, which are found to be formed by a few traders, implying potential intention of price manipulation. We further investigate the dynamics of volatility, trading volume, average trade size and turnover around the transactions associated with the abnormal motifs for large, medium and small trades. It is found that these variables peak at the abnormal events and exhibit a power-law accumulation in the pre-event time period and a power-law relaxation in the post-event period. We also find that the cumulative excess returns are significantly positive after buyer-initiated suspicious trades and exhibit a mild price reversal after seller-initiated suspicious trades. These findings can be better understood in favor of price manipulation. Our work shed new lights into the detection of price manipulation resorting to the abnormal motifs of complex trading networks.
\begin{keywords}
  Stock trading network; order flow; abnormal motif; price manipulation
\end{keywords}
\end{abstract}
\vspace{12pt}

\section{Introduction}

Price manipulation is a ubiquitous phenomenon in equity markets \citep{Putnins-2012-JES}. \cite{Cumming-Johan-2008-ALER} composes a rich list of manipulation techniques and \cite{Putnins-2012-JES} provides a taxonomy of manipulation techniques. In general, there are three categories of price manipulations: action-based manipulation, information-based manipulation and trade-based manipulation \citep{Allen-Gale-1992-RFS}. Action-based manipulation is based on actions that change the actual or perceived value of the asset, information-based manipulation is implemented by releasing false information or spreading false rumors, and trade-based manipulation is based purely on buying and selling securities without taking any publicly observable actions or spreading false information \citep{Allen-Gale-1992-RFS}. Unlike the first two categories, trade-based manipulation is easier to conduct and thus more common, and it is possible to design statistical approaches for the detection of trade-based manipulations.

Empirical studies have been conducted at the macro and micro view angles. At the macro level, researchers investigate the behavior of stock price time series trying to identify abnormality. \cite{Felixson-Pelli-1999-JMFM} and \cite{Hillion-Suominen-2004-JFinM} study the closing price manipulation, and \cite{ComertonForde-Putnins-2011-JFI} propose an index of the probability and intensity of closing price manipulation. \cite{Mahoney-1999-JFE} and \cite{Jiang-Mahoney-Mei-2005-JFE} study the alleged stock pools of the 1920s through abnormal turnover and returns and find evidence of informed trading rather than manipulation. \cite{Ogut-Doganay-Aktas-2009-ESA} and \cite{Diaz-Theodoulidis-Sampaio-2011-ESA} design algorithms for the detection of stock-price manipulation based on data mining techniques.

Recently, owning to the availability of order flow data, there are also studies from the microscopic angle. \cite{Sun-Cheng-Shen-Wang-2010-PP} investigate the distributions of the transaction number and trading volume of individual traders trading the same stock and find that the distributions of 45 non-manipulated stocks exhibit nice power-law tails, while the distributions for 7 manipulated stocks do not have power-law tails and have a obvious hump. \cite{Sun-Cheng-Shen-Wang-2011-PA} further investigate 100 non-manipulated stocks and 8 manipulated stocks. They uncover that the seller-buyer ratio is strongly correlated with the stock return for non-manipulated stocks, while the correlation is marginal for manipulated stocks. \cite{Sun-Shen-Cheng-Wang-2012-PLoS1} study stock complex trading networks and find that manipulated stocks have higher degree-strength correlations than non-manipulated stocks. These techniques have an important potential for the detection of manipulated stocks. Following the same idea, \cite{Sun-Shen-Cheng-Wang-2012-PLoS1} investigate the ratios of strength to degree for individual traders and are able to identify abnormal traders as candidates of manipulating traders.

There are also studies aiming at detecting collusive cliques or pools based on trading networks. To identify cliques that are defined as bidirectional complete subgraphs, \cite{Palshikar-Apte-2008-DMKD} adopt graph clustering methods including the shared nearest neighbors algorithm of \cite{Jarvis-Patrick-1973-IEEEtc}, the mutual nearest neighbors algorithm of \cite{Gowda-Krishna-1978-PtnR} and a new collusion clustering algorithm. They find that the two latter algorithms perform well. \cite{Islam-Haque-Alam-Tarikuzzaman-2009-ICCIT} propose a Markov clustering algorithm that is able to successfully detect circular trading, which cannot be identified using the methods in \cite{Palshikar-Apte-2008-DMKD}. In addition, \cite{Franke-GeyerSchulz-Hoser-2006-bk} and \cite{Franke-Hoser-Schroder-2008-bk} apply spectral analysis of the trading networks of traders in a prediction market and successfully identify clusters of manipulators.

Alternatively, \cite{Wang-Zhou-Guan-2012-Nc} use the similarity of the trading activities among investors to detect candidate collusive cliques. In doing so, they construct the time series of aggregated order volumes calculated in given time intervals of individual traders and calculate the correlation matrix. When a proper threshold of correlation coefficient is set, the candidate collusive cliques of traders can be identified. This method is reminiscent of the works to classify trader clusters using random matrix theory on the inventory variation time series \citep{Lillo-Moro-Vaglica-Mantegna-2008-NJP,Zhou-Mu-Kertesz-2012-NJP} or statistically validated networks of trading activity \cite{Tumminello-Lillo-Piilo-Mantegna-2012-NJP}.

In this work, we report preliminary results on the behaviors of potential manipulating traders identified from the complex trading networks. For each stock, we can construct a trading network, where the traders are nodes and a directed link is assigned from one trader to another if the former sells some shares to the latter. We investigate the properties of three specific motifs that correspond to possible manipulation techniques. We find that these abnormal motifs do contain rich information. The underlying idea of this Letter is obviously different from those in \cite{Palshikar-Apte-2008-DMKD} and \cite{Islam-Haque-Alam-Tarikuzzaman-2009-ICCIT}, who also work on trading networks of traders.

The study of security trading networks (STNs) has a long history and can be at least traced back to \cite{Baker-1981-PhD,Baker-1984-AJS}. However, the progress is very slow. The situation changes gradually in recent years since trading data are easier to obtain and network science flourishes. \cite{Franke-GeyerSchulz-Hoser-2006-bk} and \cite{Franke-Hoser-Schroder-2008-bk} analyze a prediction market to detect clusters of price manipulators. \cite{Wang-Tseng-Tai-Lai-Wu-Chen-Li-2008-EPJB} study the network topology of a prediction market which is an experimental futures exchange. \cite{Tseng-Li-Chen-Wang-2009-ACS} propose an agent-based model with ``zero-intelligence'' traders under the continuous double auction market mechanism to explain the power-law degree distribution of the trading network. \cite{Tseng-Li-Wang-2010-EPJB} investigate the trading network of a prediction market to study the dynamics of wealth accumulation. \cite{Tseng-Lin-Lin-Wang-Li-2010-PA} study the statistical properties of a trading network in an agent-based double auction model.

There are also several studies focusing on real security markets. \cite{Adamic-Brunetti-Harris-Kirilenko-2012-SSRN} construct trading networks using transaction-level data for the September 2009 E-mini S\&P 500 futures contract and find that network metrics are highly contemporaneously correlated with returns, volatility, volume, duration, and market liquidity and strongly Granger-cause intertrade duration and trading volume. \cite{Jiang-Zhou-2010-PA} explore the statistical properties of trading networks of a highly liquid stock traded on the Shenzhen Stock Exchange for the whole year of 2003. \cite{Wang-Zhou-Guan-2011-PA} concerns with the statistical characteristics of the trading networks of all commodities traded on the Shanghai Futures Exchange from July to September, 2008. \cite{Sun-Cheng-Shen-Wang-2011-PA} and \cite{Sun-Shen-Cheng-Wang-2012-PLoS1} analyze the trading network of manipulated and non-manipulated stocks listed on the Shanghai Stock Exchange and report significant differences in the topological properties between the two categories of stocks.

\section{Basic description}

\subsection{Data sets}

The data sets under investigation are the history records of order flows of 43 stocks traded on the Shenzhen Stock Exchange during the whole year of 2003, including 32 A-shares and 11 B-shares. The list of codes for the A-share stocks are 000001, 000002, 000009, 000012, 000016, 000021, 000024, 000027, 000063, 000066, 000088, 000089, 000406, 000429, 000488, 000539, 000541, 000550, 000581, 000625, 000709, 000720, 000778, 000800, 000825, 000839, 000858, 000898, 000917, 000932, 000956 and 000983, and the list of codes for the 11 B-share stocks are 200002, 200012, 200016, 200024, 200429, 200488, 200539, 200541, 200550, 200581 and 200625. The A-share stocks are constituents of the Shenzhen Stock Exchange Component Index and the B-shares are associated with some of the A-share stocks.

Each submitted or canceled order contains the following information: time stamp of order placement or cancelation, unique encrypted identity of the trader, limit order price, order size, and buy-sell identifier (buy, sell or cancelation). We are thus able to reconstruct the limit order book and trace the transactions. In this work, only transactions in the continuous double auction are considered. For more information about the Shenzhen stock market, see \cite{Mu-Zhou-Chen-Kertesz-2010-NJP} and \cite{Zhou-Mu-Kertesz-2012-NJP}.

\subsection{Construction of stock trading networks}

We construct an entire trading network for each stock. The trading network construction approach is the same as in \cite{Jiang-Zhou-2010-PA} and \cite{Sun-Shen-Cheng-Wang-2012-PLoS1}. Each trader who bought or sold the stock enters the network as a node. A directed link is formed between two traders if they had transactions and the direction of the link is from the seller to the buyer. When a trader places an effective market order, it is possible that the order is executed by several orders on the limit order book submitted by different traders. In this case, the local network structure is a star-like graph with the central node acting as a source if the trader sells or a sink if the trader buys.

\subsection{Determination of abnormal trading motifs}

\begin{figure}
  \centering
  \begin{minipage}{50mm}
    \resizebox*{4cm}{!}{\includegraphics{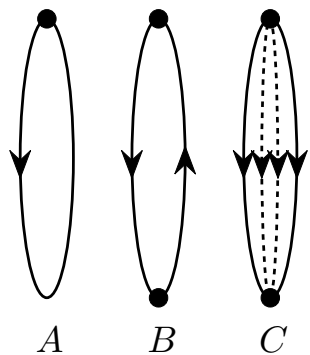}}%
    \label{Fig:AbnormalMotifs}
  \end{minipage}
  \caption{Graphical illustration of abnormal trading motifs: (A) Self-loop, (B) Two-node loop, (C) Two-node multiple arcs.}
\end{figure}

The constructed trading networks record the patterns of order execution in limit order book and of the flows of cash and stock shares among investors, which provides a potential opportunity to detect market manipulations of some traders and to further investigate their influences on the market's behaviors. By scanning all the trading networks, we find that there are some motif patterns in contrast with the intuitions, which can be considered as evidence in favor of market manipulations of some investors. Network motifs are building blocks of complex networks \citep{Milo-ShenOrr-Itzkovitz-Kashtan-Chklovskii-Alon-2002-Science,Milo-Itzkovitz-Kashtan-Levitt-ShenOrr-Ayzenshtat-Sheffer-Alon-2004-Science}, and motifs evolve in time-varying networks \citep{Kovanen-Karsai-Kaski-Kertesz-Saramaki-2011-JSM}. When three-node motifs are considered in directed networks, there are 13 possible motifs without self-loops \citep{Milo-ShenOrr-Itzkovitz-Kashtan-Chklovskii-Alon-2002-Science}. In this work, we investigate three motifs as depicted in Figure \ref{Fig:AbnormalMotifs}. The determination of motifs is based on economic considerations.

%\begin{figure}
%  \centering
%  \includegraphics[width=4cm]{Fig_STNMotif_Illustration_AbnormalMotifs.eps}
%  \caption{\label{Fig:AbnormalMotifs0} Graphical illustration of abnormal trading motifs: (A) Self-loop, (B) Two-node loop, (C) Two-node Multiple arcs.}
%\end{figure}

Motif $A$ in figure \ref{Fig:AbnormalMotifs} is a self-loop, containing a single trader who sells shares to herself. Such motifs are reminiscent of wash sales, which are improper transactions in which the buyer and seller is the same person such that there is no genuine change in ownership \citep{Putnins-2012-JES}. If a trader utilizes the technique of wash sales, her trading behavior will be identified as motif $A$ from the trading network. On the other hand, a motif $A$ trader is probably not a manipulator, but the probability is very low.

Motif $B$ in figure \ref{Fig:AbnormalMotifs} is a two-node loop, in which two traders (or more precisely two stock accounts) exchange shares of the same stock. Motif $B$ is the simplest structure embedded in the manipulation technique of stock pools. As described by \cite{Putnins-2012-JES}, stock pools are a collusive group of manipulators trading shares back and forth among themselves. Cliques investigated in \cite{Palshikar-Apte-2008-DMKD} or collusion sets of circular trading investigated in \cite{Islam-Haque-Alam-Tarikuzzaman-2009-ICCIT} are special examples of stock pools. Motif $B$ can be observed not only in cliques and some circular trading sets but also in other pools.

Motif $C$ in figure \ref{Fig:AbnormalMotifs} contains multiple (at least two) links with the same direction, which happens when one account repeatedly sells to or buys from the same counterparty. The occurrence of network structure of motif $C$ is more common than motif $A$ and motif $B$. Motif $C$ might correspond to a normal behavior when a trader buys or sells the same stock several times and encounters the same counterparty by coincidence. On the other hand, motif $C$ might also appear between two traders within a same pool.

\section{Topological properties of abnormal trading motifs}

\begin{figure}
  \centering
  \includegraphics[width=5cm]{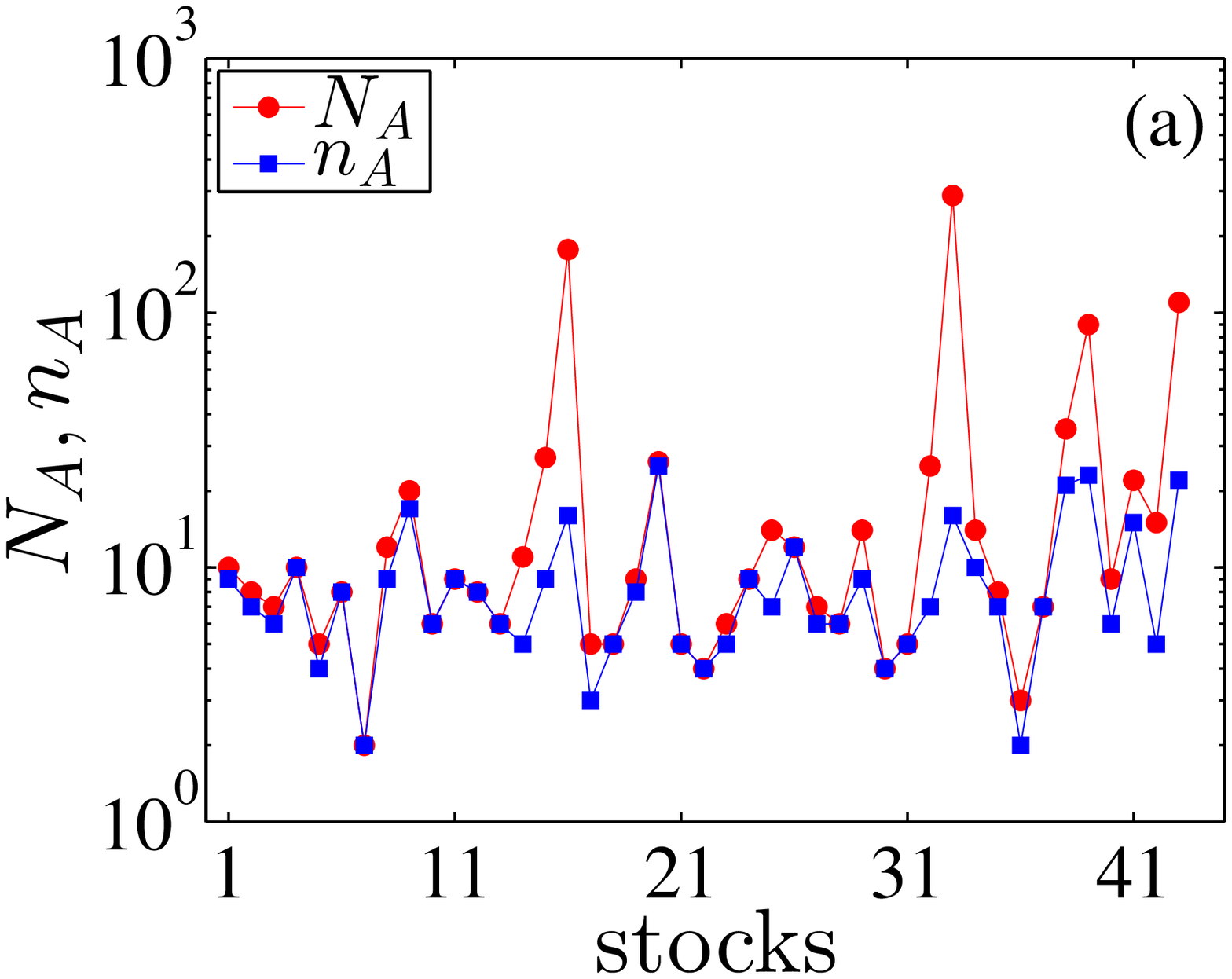}
  \includegraphics[width=5cm]{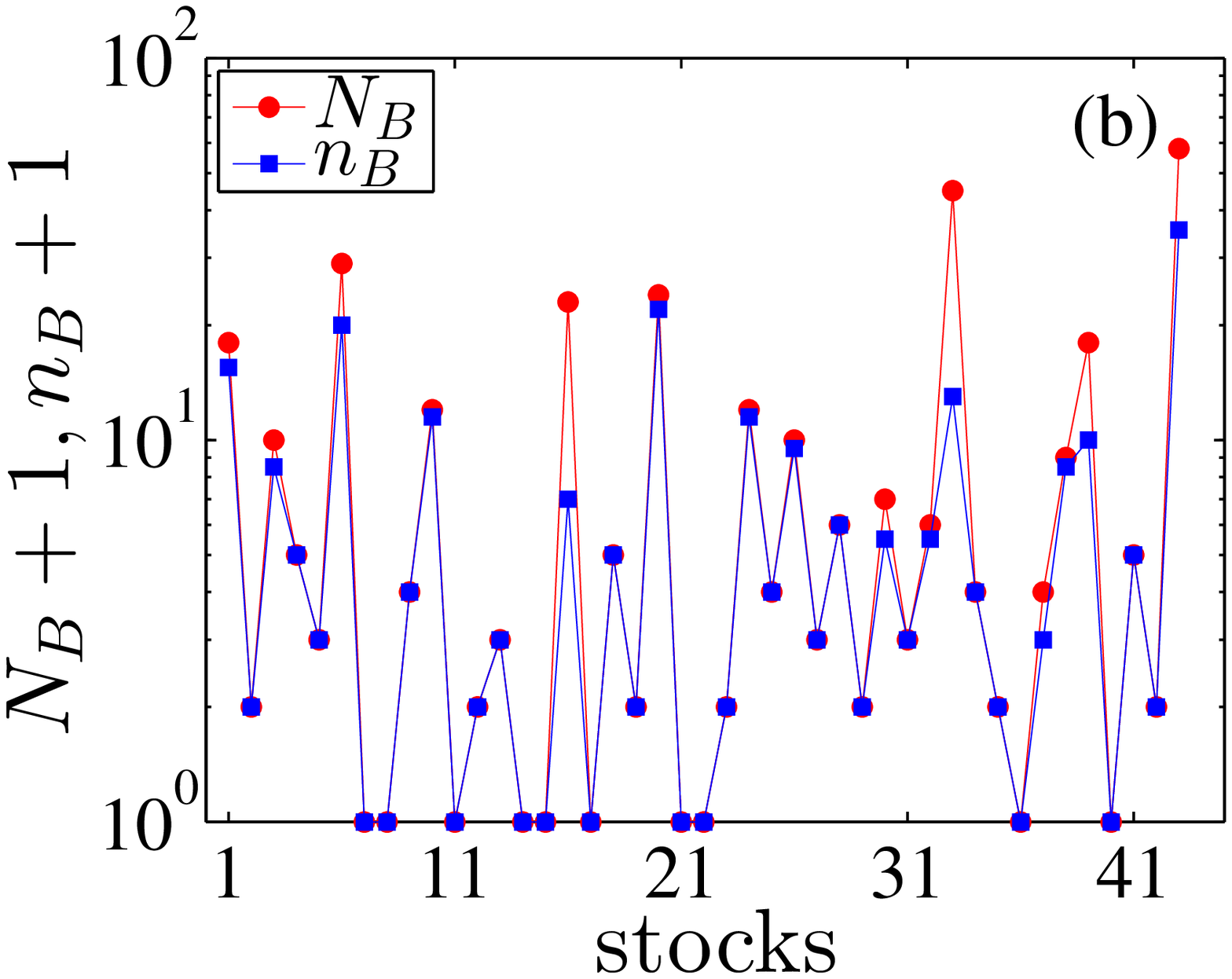}
  \includegraphics[width=5cm]{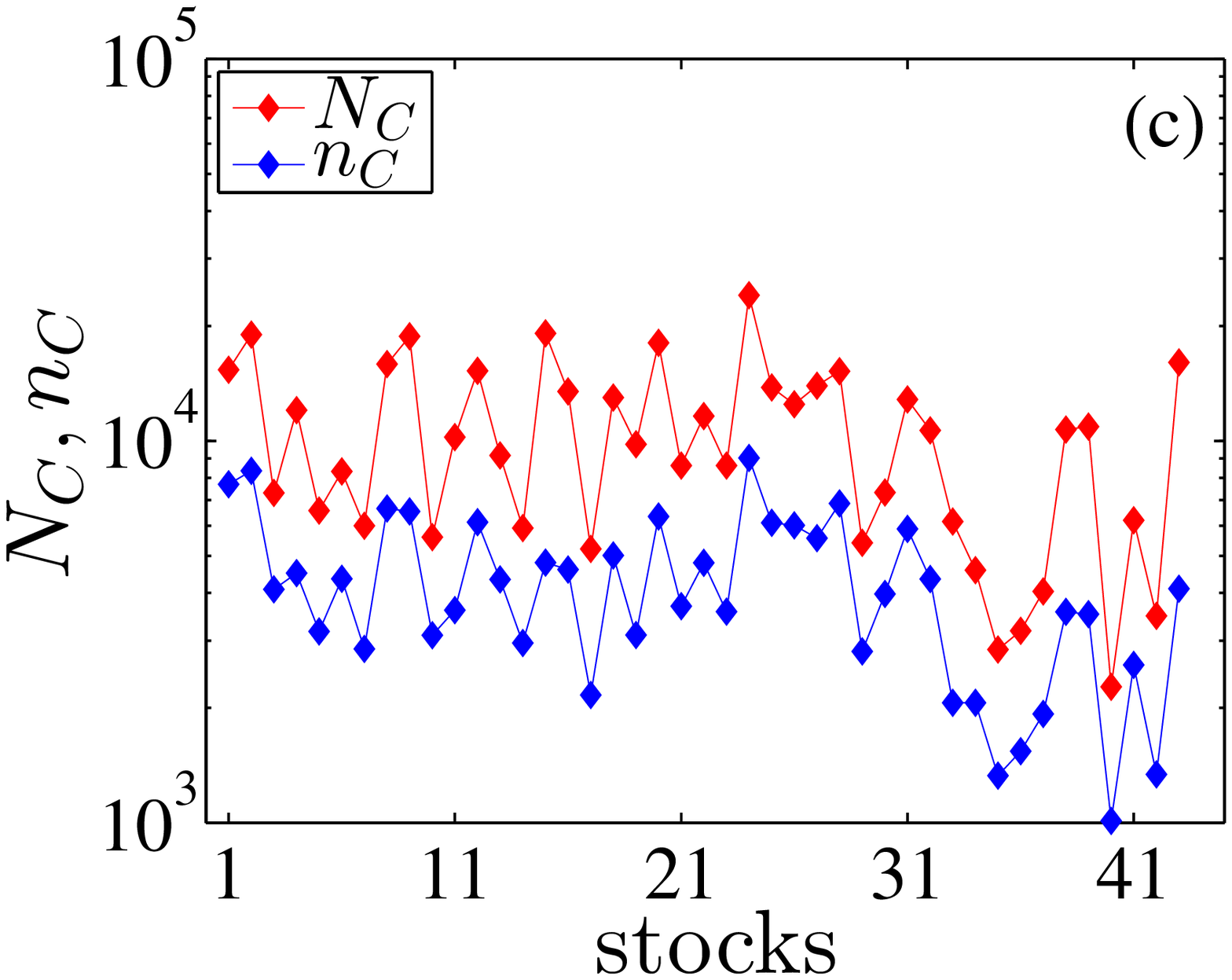}
  \caption{\label{Fig:AbnormalMotifs:Num} The occurrence numbers of motifs $A$, $B$ and $C$ ($N_A$, $N_B$ and $N_C$) and the associated
  characteristic numbers of traders ($n_A$, $n_B$ and $n_C$) for different stocks. Each point corresponds to a stock.}
\end{figure}

For each stock, we have identified all motifs of the three types as depicted in figure \ref{Fig:AbnormalMotifs}. The occurrence numbers of different motifs identified are plotted in figure \ref{Fig:AbnormalMotifs:Num}. In each plot, we also show the characteristic numbers of traders for each stock. Specifically, $n_A$, $2n_B$ and $2n_C$ are the numbers of traders exhibiting respectively motifs $A$, $B$ and $C$. Hence $n_M<N_M$ if there are overlapping $M$-motifs (or at least two motifs share a same trader) for a stock, where $M=A$, $B$, or $C$. Otherwise, we have $n_M=N_M$.

Figure \ref{Fig:AbnormalMotifs:Num}(a) shows that $n_A=N_A$ for 16 stocks whose motifs do not have multiple links in the self-loops and $n_A<N_A$ for other 27 stocks. Figure \ref{Fig:AbnormalMotifs:Num}(b) shows that there are 10 stocks that do not have motif $B$ since $N_B=0$, 18 stocks stocks have isolated $B$-motifs since $n_B=N_B$, and 15 stocks have overlapping $B$-motifs since $n_B<N_B$. Figure \ref{Fig:AbnormalMotifs:Num}(c) shows that all stocks have $C$-motifs and many of these motifs overlap since $n_C \ll N_C$. The observation of $n_M{\ll}N_M$ for some or all stocks provides further evidence of possible price manipulation.

\begin{figure}
  \begin{center}
  \begin{minipage}{140mm}
    \subfigure[]{\resizebox*{7cm}{!}{\includegraphics{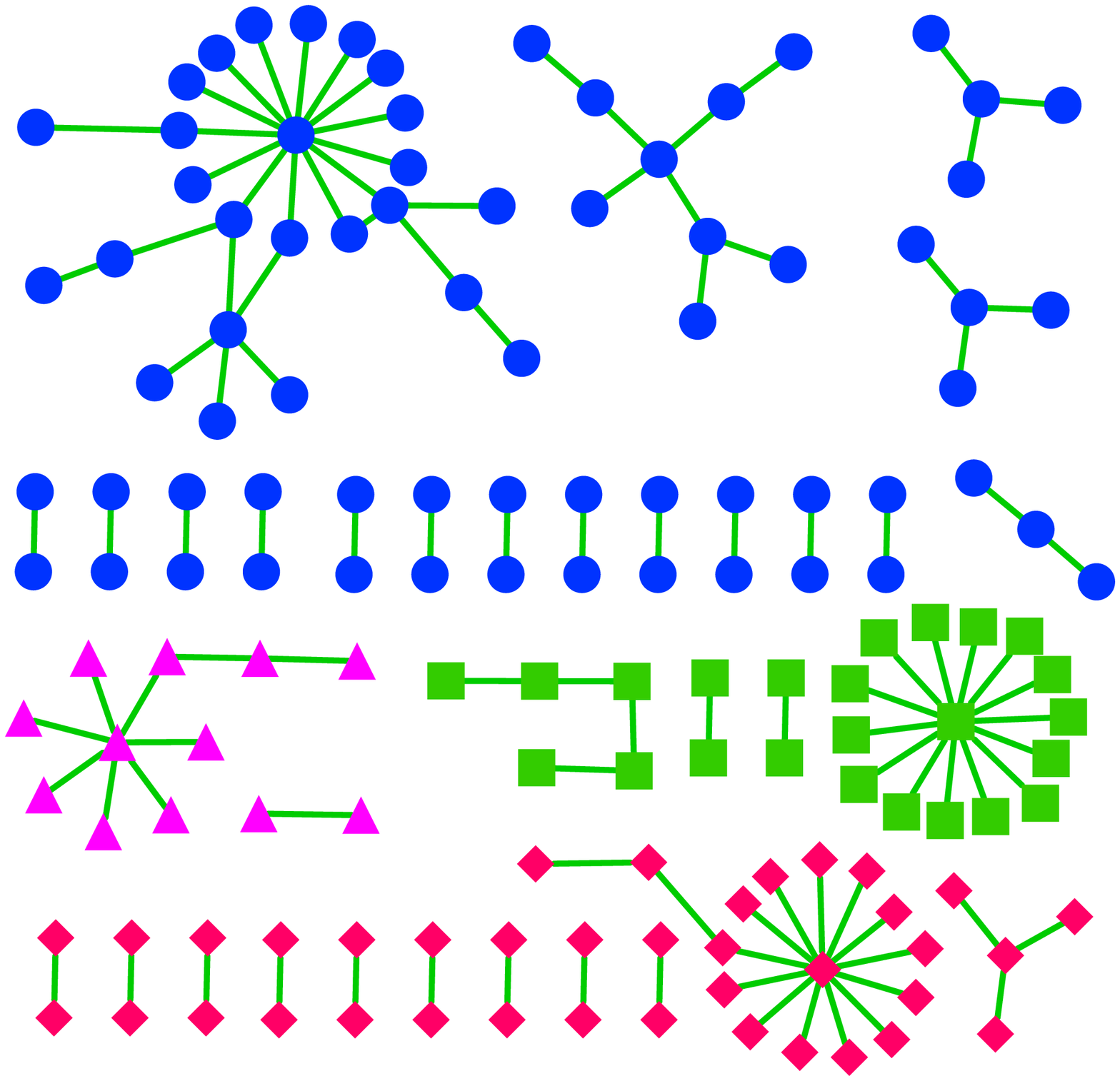}}\label{Fig:AbnormalMotifs:B}}%
    \subfigure[]{\resizebox*{7cm}{!}{\includegraphics{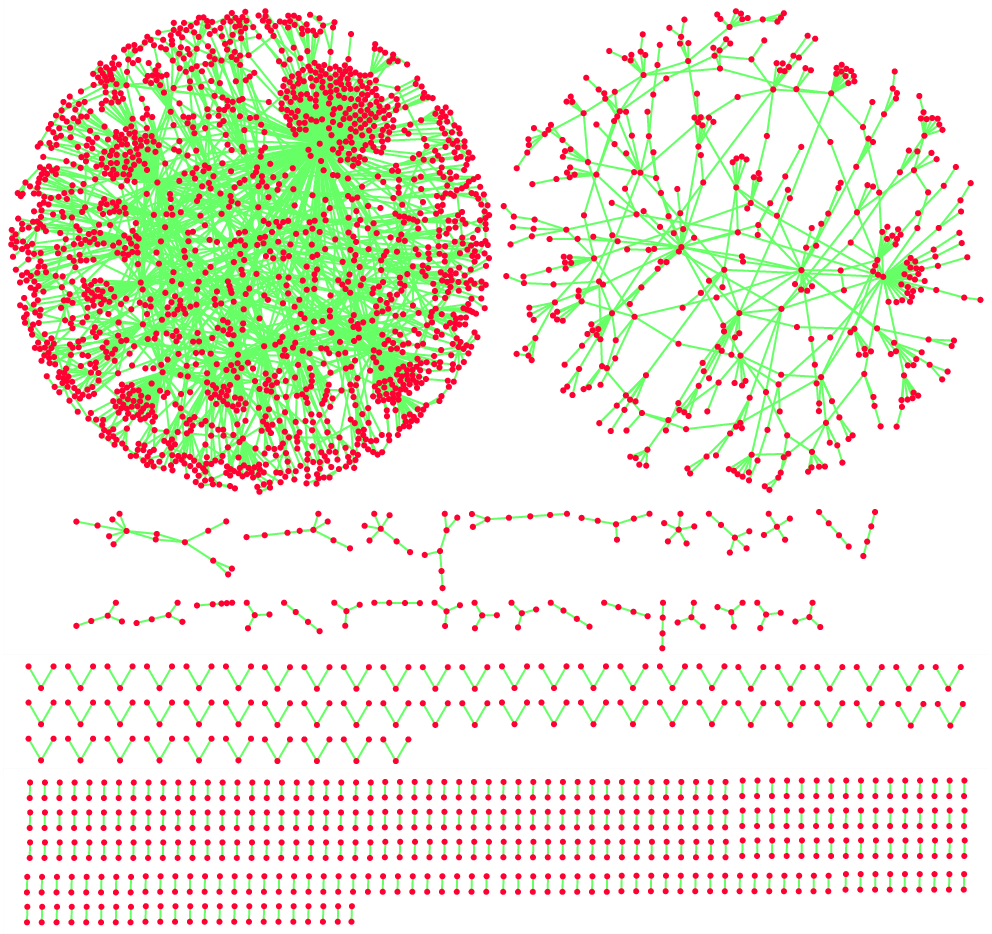}}\label{Fig:AbnormalMotifs:C}}%
    \label{Fig:AbnormalMotifs:B:C}
  \end{minipage}
  \end{center}
  \caption{(a) The entire networks containing all nodes and all links in the identified $B$-motifs for four stocks: 000021 ({\color{red}{$\blacklozenge$}}), 000539 ({\small\color{magenta}{$\blacktriangle$}}), 200002 ({\tiny\color{green}{$\blacksquare$}}), 200625 ({\large\color{blue}{$\bullet$}}). Each edge in the plot represents a pair of bidirectional links. (b) The entire network containing all nodes and all links in the identified $C$-motifs for stocks 200024. Each edge stands for multiple directed links.}
\end{figure}

To illustrate the complex structure of each motif type, we show in figure \ref{Fig:AbnormalMotifs:B:C}(a) the entire networks containing all $B$-motifs for four stocks. We find that most connected clusters have tree-like or star-like structures. Only one stock (200625) possesses a trading circle of four traders and no stock has cliques or other loops. Figure \ref{Fig:AbnormalMotifs:B:C}(b) illustrates the entire network containing all nodes and links in the identified $C$-motifs of a typical stock. We observe that there are two large sub-graphs and many small-size clusters including a lot of isolated $C$-motifs and three-node chains. There are also loops in some clusters. We note that the size distribution of the clusters of all stocks for motif $C$ has a nice power law behavior:
\begin{equation}
  p(S) \sim S^{-(\alpha+1)},
\end{equation}
where the tail exponent is estimated to be $\alpha\approx3.5$.

%
%\begin{figure}
%  \centering
%  \includegraphics[width=7cm]{Fig3_Clusters_MotifB.eps}
%  \includegraphics[width=7cm]{Fig3_Clusters_MotifC.eps}
%  \caption{\label{Fig:AbnormalMotifs:B:C} (a) The entire networks containing all nodes and all links in the identified $B$-motifs for four stocks: 000021 ({\color{red}{$\blacklozenge$}}), 000539 ({\small\color{magenta}{$\blacktriangle$}}), 200002 ({\tiny\color{green}{$\blacksquare$}}), 200625 ({\large\color{blue}{$\bullet$}}). Each edge in the plot represents a pair of bidirectional links. (b) The entire network containing all nodes and all links in the identified $C$-motifs for stocks 200024. Each edge stands for multiple directed links.}
%\end{figure}

\begin{figure}
  \centering
  \includegraphics[width=6.5cm]{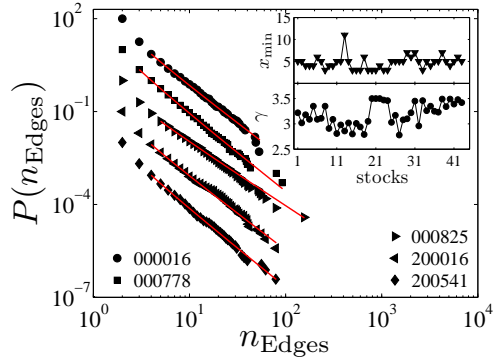}
  \caption{\label{Fig:SA:AbnormalMotifs:CDF:MotifC} The cumulative probability distribution of the number edges in motif $C$ for five random
  chosen stocks. The data points of 000016, 000778, 200016, and 200541 are translated vertically by a factor of 100, 10, 0.1, and 0.01 for
  better visibility, respectively. The straight lines give the best fits to the data.}
\end{figure}

By definition, motif $A$ is composed of only one edge and motif $B$ is constituted by two edges. In contrast, a $C$-motif may contain double, trinary, quadruple, and more edges. Therefore, it is interesting to explore the distribution of the number of edges $n_{\rm{Edges}}$ in each $C$-motif for each stock. Figure~\ref{Fig:SA:AbnormalMotifs:CDF:MotifC} plots the empirical cumulative distribution of $n_{\rm{Edges}}$ for five random chosen stocks. It is found that there are $C$-motifs which have more than 100 edges and the cumulative distributions have power-law tails:
\begin{equation}
  P(n_{\rm{Edges}}) \sim n_{\rm{Edges}}^{-\gamma}.
\end{equation}
Following \cite{Clauset-Shalizi-Newman-2009-SIAMR}, we can obtain the estimates of the power-law exponent $\gamma$ and the low boundary $x_{\min}$ of the power-law behaviors. The inset of figure~\ref{Fig:SA:AbnormalMotifs:CDF:MotifC} presents the values of $\gamma$ and $x_{\min}$ for different stocks. The average value of the power-law exponents for the whole sample is $\langle \gamma \rangle = 3.19 \pm 0.23$.

\section{Market dynamics surrounding the abnormal trading edges}

We investigate in this section the dynamics of four commonly used financial quantities before and after abnormal transactions associated with the identified motifs, including volatility $v$ determined as the absolute return of mid-price, trading volume $\omega_{\rm{cum}}$ defined as the total trade size in each minute, average trade size $\omega_{\rm{ave}}$ defined as the trading volume divided by the number of trades in each minute, and turnover value $f$ defined as the sum of transaction values (product of trade size and the corresponding transaction price) in each minute. All these variables exhibit pronounced intraday patterns. Consider a given financial variable $x(d, t')$, where $d$ identifies the trading days and $t' = 1, 2, \cdots, 240$ are 1-min intervals within each trading day. The intraday pattern is determined as follows \citep{Jiang-Chen-Zhou-2009-PA}:
\begin{equation}
  {\bar{x}}(t')=\frac{1}{N_d}\sum_{d=1}^{N_d}x(d, t'),
  \label{Eq:IntPtn}
\end{equation}
where $N_d$ is the number of trading days for the stock. We remove the intraday patterns before aggregating the events from different stocks:
\begin{equation}
  x_{r}(d,t') = x(d,t') / {\bar{x}}(t').
  \label{Eq:RemIntPat}
\end{equation}

To investigate the market dynamics around the detected abnormal trading motifs (or manipulation edges), we employ the approach commonly used for analyzing market reactions to public news or announcements \citep{Fleming-Remolona-1997-EPR,Fleming-Remolona-1999-JF,Joulin-Lefevre-Grunberg-Bouchaud-2008-Wilmott,Erenburg-Lasser-2009-RFE,GrossKlussmann-Hautsch-2009-SSRN} and for understanding market behaviors around predefined extreme price changes \citep{Zawadowski-Kertesz-Andor-2004-PA,Hamelink-2003-JFc,Zawadowski-Andor-Kertesz-2006-QF,Toth-Kertesz-Farmer-2009-EPJB,Ammann-Kessler-2009-AFE,Ponzi-Lillo-Mantegna-2009-PRE,Mu-Zhou-Chen-Kertesz-2010-NJP}. Each directed link in the identified motifs stands for a transaction and is called an event in this work. We categorize the events into three groups according to the trade sizes of events. For each stock, we sort the events ensuring that the associated trade sizes are in the decreasing order. Twenty events are picked out into the large-size group in a iterative way. The first event with the largest trade size is chosen. The second event is chosen only if it does not occur on the same trading day as the first event. The third event is chosen only if it does not occur on the same trading days of previously chosen events. This procedure repeats until 20 events are chosen. The median-size group and the small-size group are determined in a similar way, ensuring that the events are around the median trade size and the smallest trade size, respectively. The chosen events in the large-size (median-size or small-size) groups of all stocks are put together to form the final large-size (median-size or small-size) group. We divide the trade size of each event in the three groups by the average trade size of the underlying stock and obtain the averages for the three groups which are 68.8, 0.88 and 0.0026 respectively.

We perform analysis for each group. Following the methodology of event analysis, we first determine the time point $t'_0$ when the event occurs. One can further extract the evolutionary trajectory of the financial quantity $\{x_{k\in\kappa}(t) : t = -200, \cdots, -1, 0, 1, \cdots, 200\}$ from $x_{r}(d, t')$, where $\kappa\in\{L,M,S\}$ and $t=0$ corresponds to $t'_0$. If the time to the opening or closing time is less than 200 minutes, we simply extend the time series to the previous or next trading day, which is rational since the intraday pattern has been removed \citep{Zawadowski-Kertesz-Andor-2004-PA,Mu-Zhou-Chen-Kertesz-2010-NJP}. The averages $\langle x_{\kappa}(t) \rangle$ over all events in each group can be obtained as follows:
\begin{equation}
 \bar{x}_{\kappa}(t) = \frac{1}{\|\kappa\|} \sum_{k \in \kappa} x_k(t),~~t = -200, \cdots, 200,
 \label{Eq:MinAverage}
\end{equation}
where $\|\kappa\|$ is the number of events in group $\kappa$. If there is nothing abnormal in a group, we expect to have $\bar{x}_{\kappa}(t)\approx 1$ for all $t$ values.

\begin{figure}
  \centering
  \includegraphics[width=3.6cm]{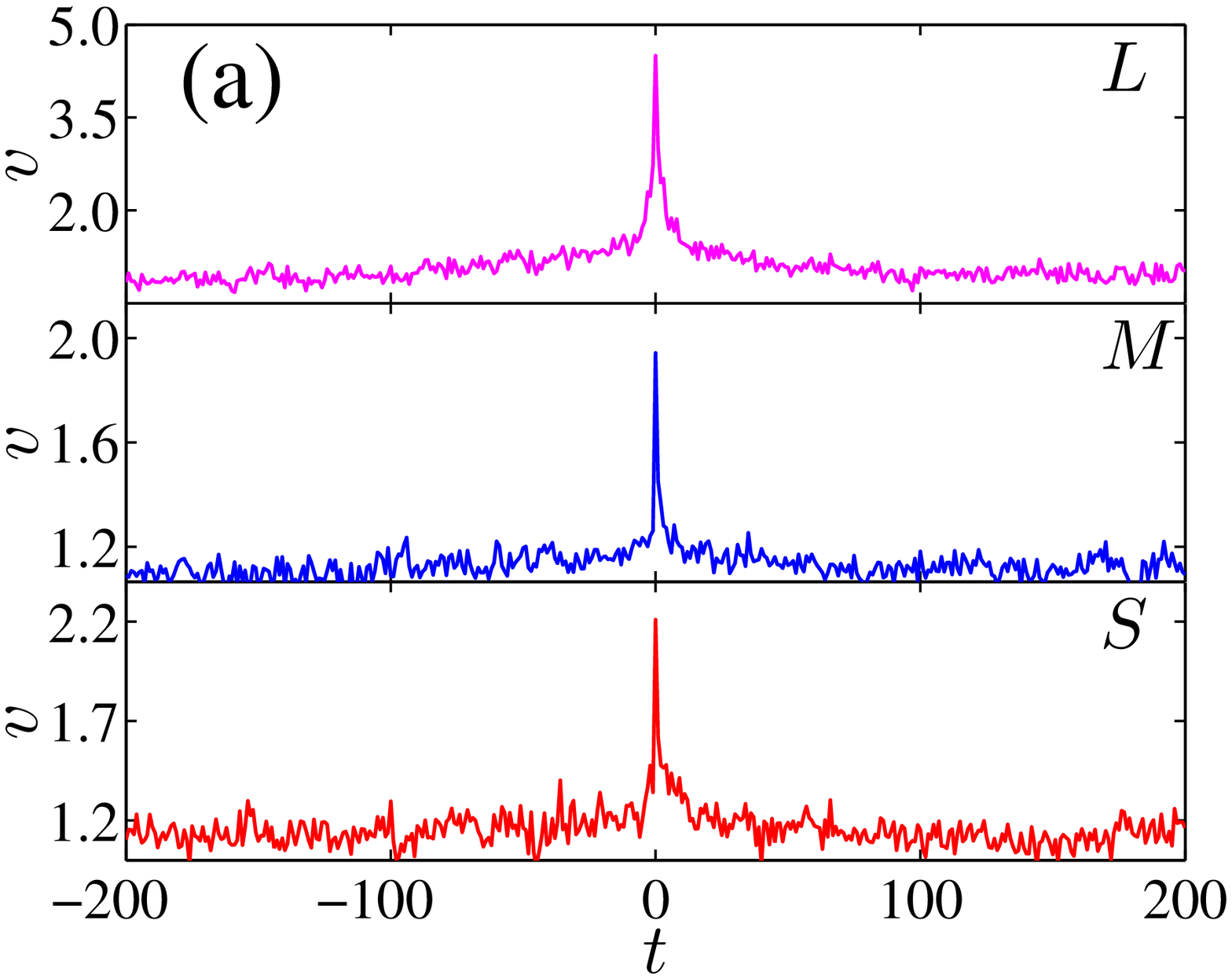}
  \includegraphics[width=3.6cm]{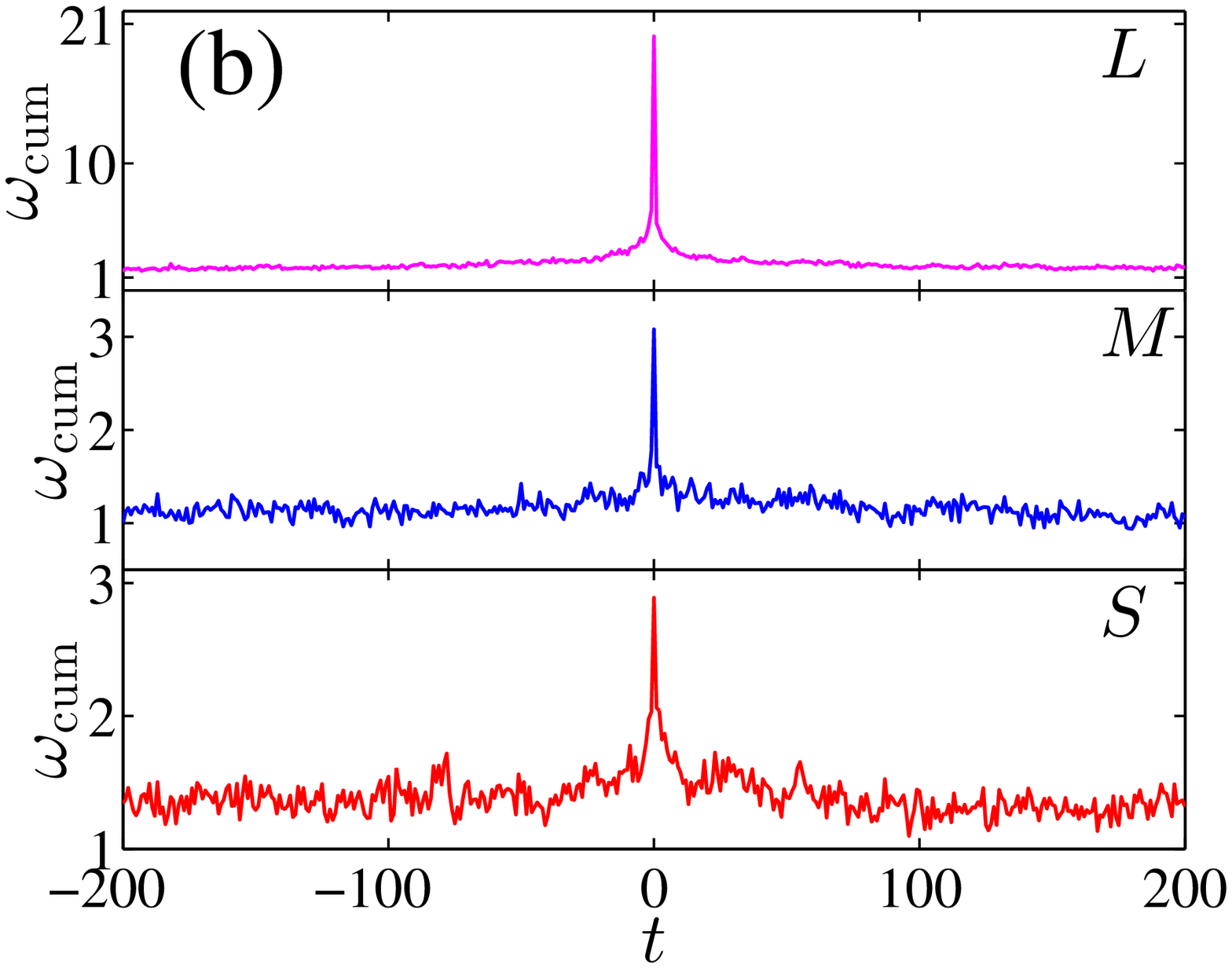}
  \includegraphics[width=3.6cm]{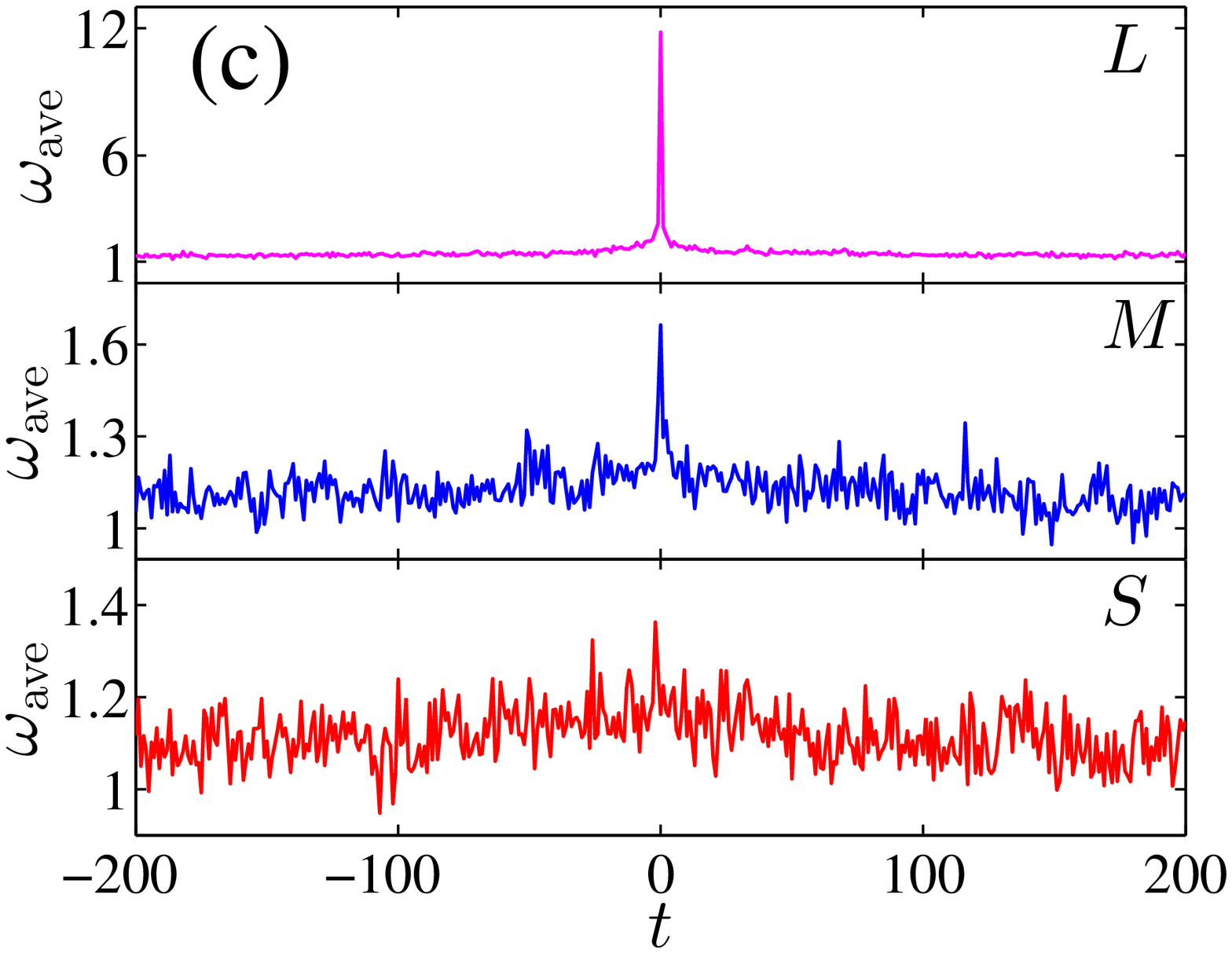}
  \includegraphics[width=3.6cm]{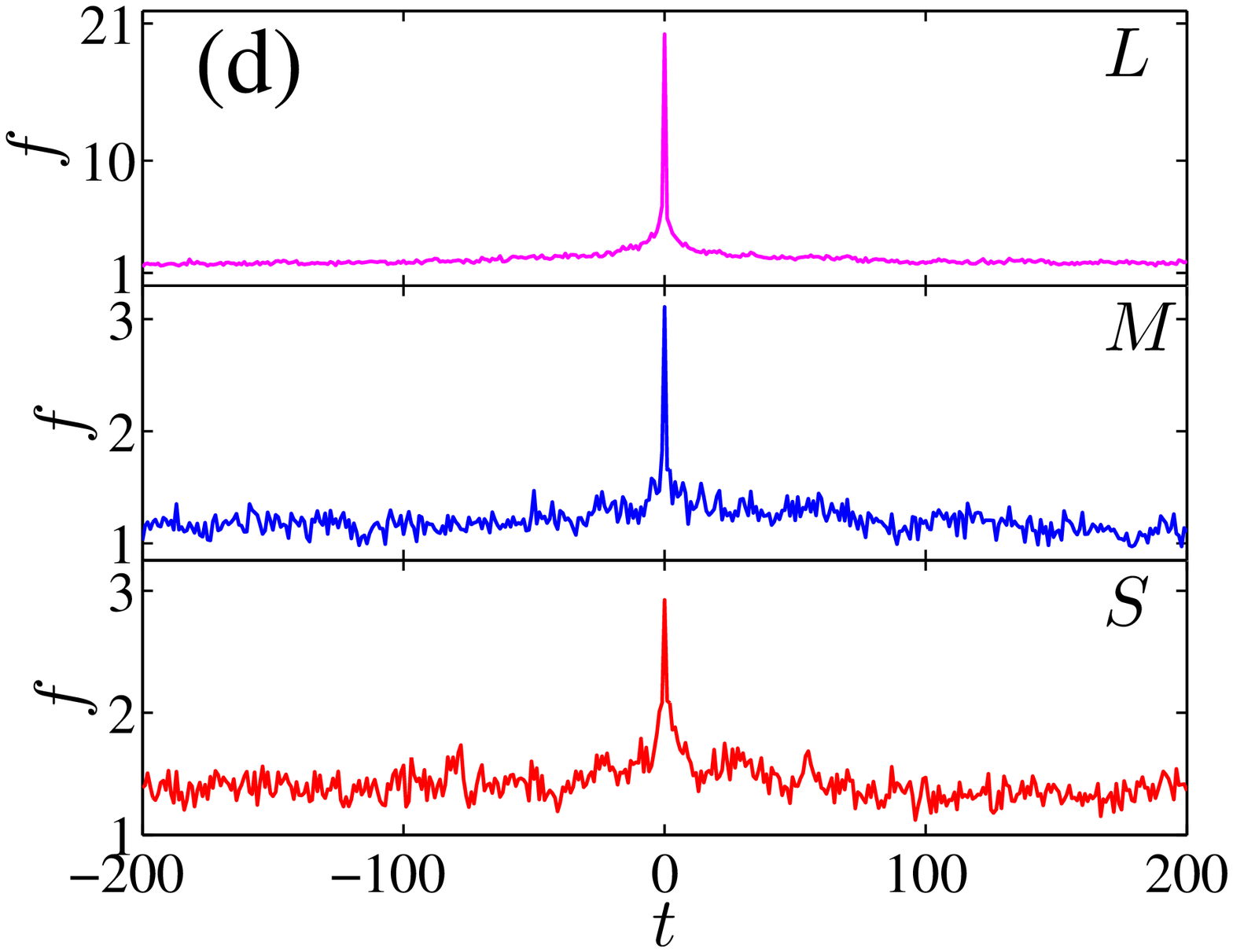}\\
  \includegraphics[width=3.6cm]{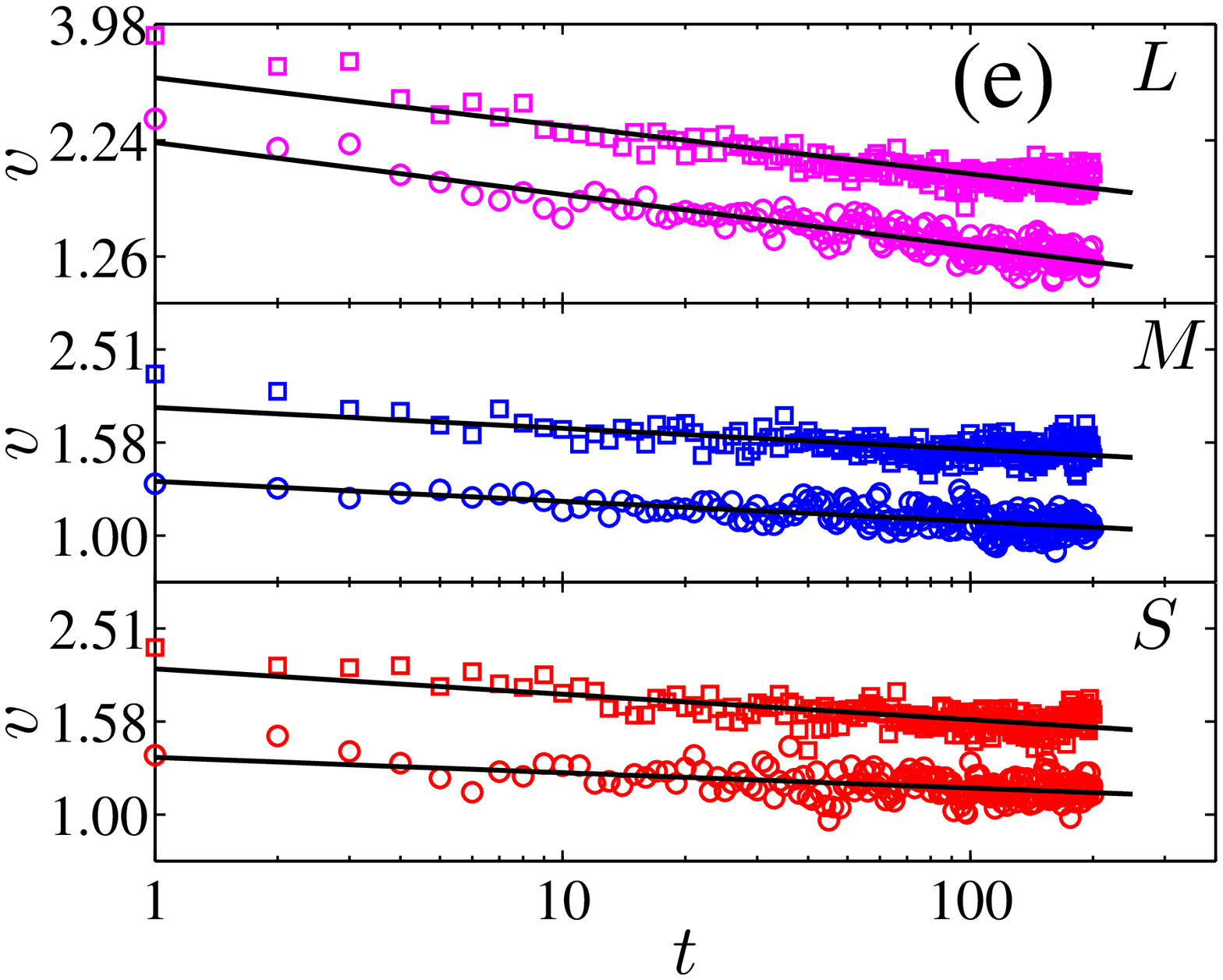}
  \includegraphics[width=3.6cm]{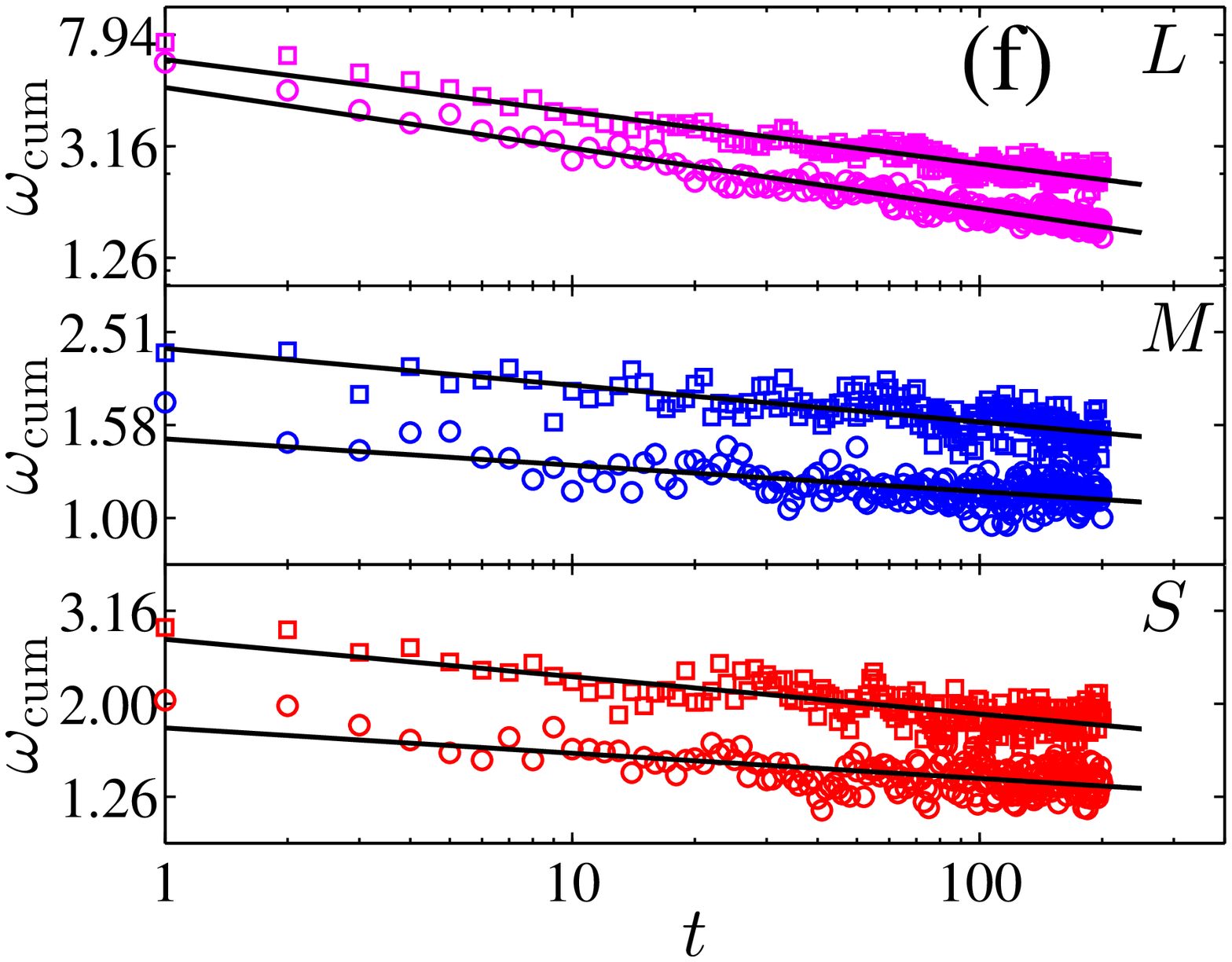}
  \includegraphics[width=3.6cm]{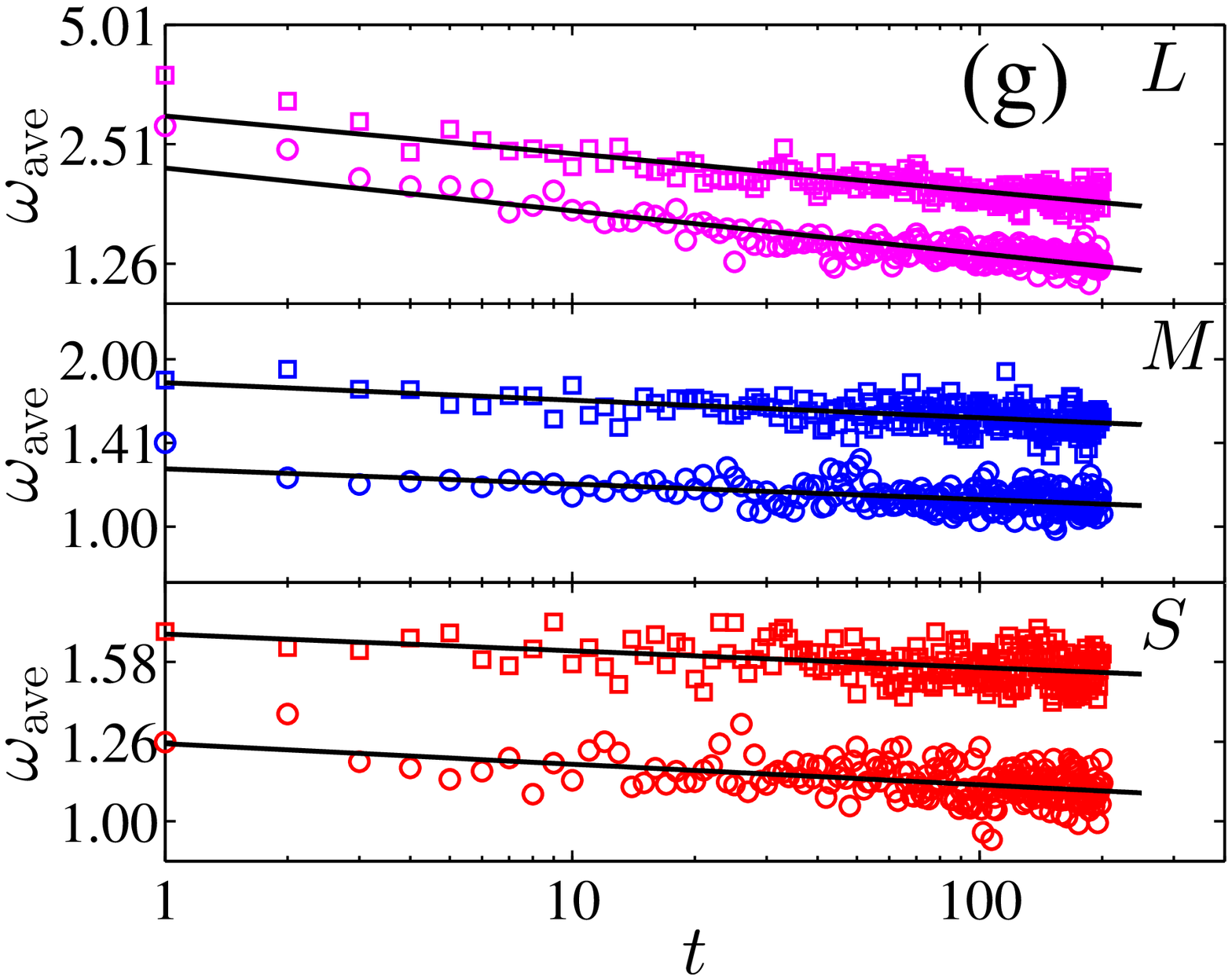}
  \includegraphics[width=3.6cm]{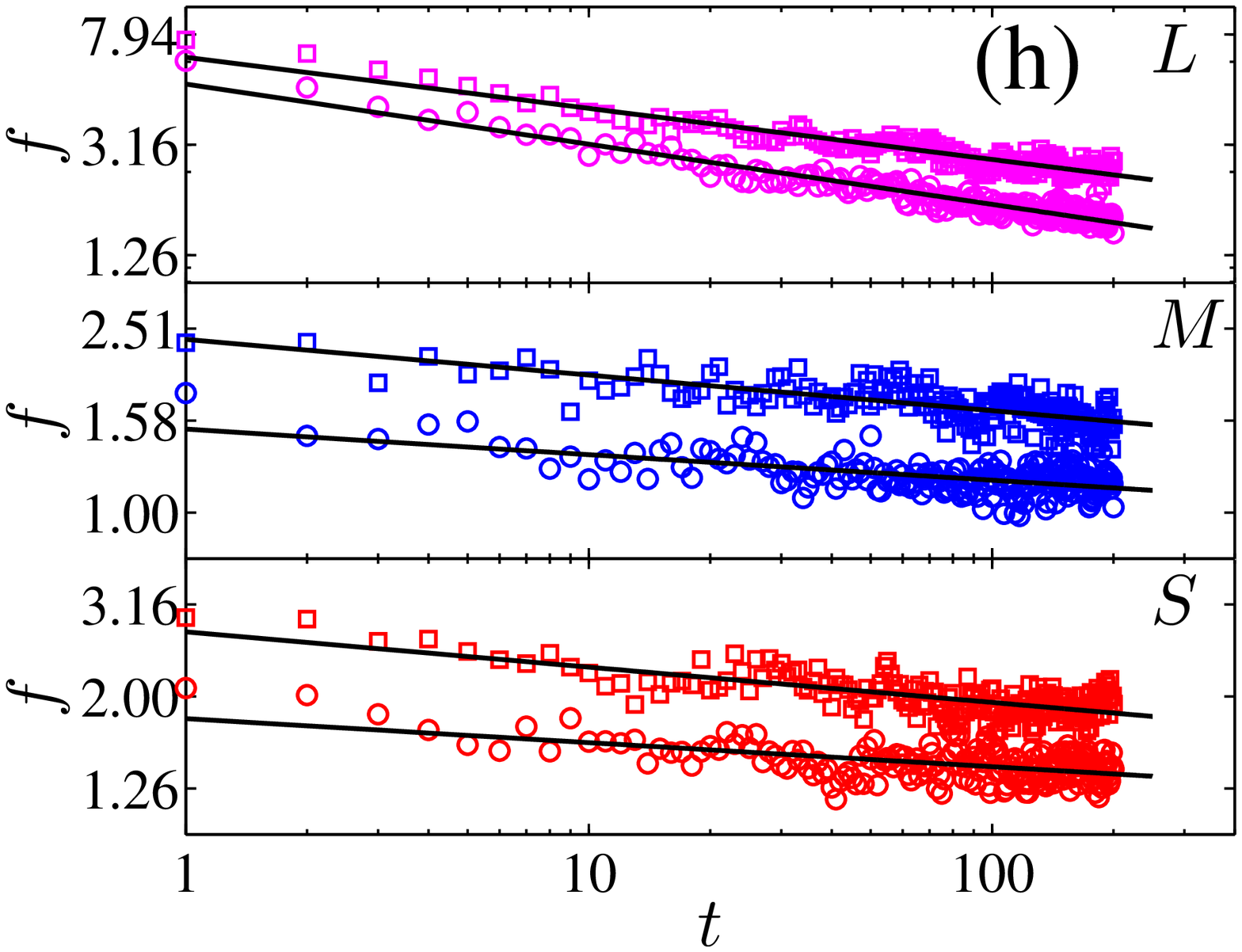}
  \caption{\label{Fig:Impact:TraFinVar} Dynamics of four representative financial variables with the intraday patterns removed around three groups ($L$, $M$, and $S$) of abnormal trading edges in linear coordinates (a-d) and logarithmal coordinates (e-f). The results for the large-size, median-size and small-size groups are labeled as $L$, $M$ and $S$ respectively. (a, e) Volatility $v$. (b, f) Trading volume per minute $\omega_{\rm{cum}}$. (c, g) Average trade size per minute $\omega_{\rm{ave}}$. (d, h) Turnover value per minute $f$.}
\end{figure}

Plots (a-d) of figure~\ref{Fig:Impact:TraFinVar} show the pre-event and post-event dynamics of four representative financial variables for the three groups. Almost all the twelve curves exhibit a marked peak at $t=0$, accompanied with a gradual accumulation when $t<0$ and a slow decay when $t>0$. We also observe that larger trade size causes more significant deviation of a financial variable from its normal value especially around the events: $\bar{x}_{L}(t)>\bar{x}_{M}(t)>\bar{x}_S(t)$. The results are qualitatively similar if we further divide the abnormal trading edges into two groups with seller-initiated trades and buyer-initiated trades.

It has been reported that the market dynamics surrounding large price changes exhibit power-law accumulations and relaxations for many financial variables \citep{Zawadowski-Kertesz-Andor-2004-PA,Zawadowski-Andor-Kertesz-2006-QF,Toth-Kertesz-Farmer-2009-EPJB,Ponzi-Lillo-Mantegna-2009-PRE,Mu-Zhou-Chen-Kertesz-2010-NJP}. It is interesting to check if this feature applies to the current case. The plots (e-h) of figure~\ref{Fig:Impact:TraFinVar} shows the results in plots (a-d) in log-log scales, where the pre-event dynamics have been reflected with respect to $t=0$. All curves exhibit nice power-law behaviors:
\begin{equation}
  \bar{x}_{\kappa}(t) \sim |t|^{-\beta_{\kappa,x}}.
\end{equation}
The estimated power-law exponents using linear regression are presented in table \ref{Tb:STNmotif:PLexponent}. For pre-event dynamics, we find $\beta_{L,x}>\beta_{M,x}>\beta_{S,x}$ for all financial variables investigated. These observations can be explained as follows. For post-event dynamics, we find $\beta_{L,x}>\beta_{M,x}\approx\beta_{S,x}$. These observations can be explained as follows. If we compare the pre-event power-law exponent with the post-event power-law exponent, we find that $\beta_{L,x,{\rm{pre}}}>\beta_{L,x,{\rm{post}}}$ for all the four quantities and $\beta_{\{M,S\},x,{\rm{pre}}}<\beta_{\{M,S\},x,{\rm{post}}}$ except for $\beta_{S,\omega_{\rm{ave}}}$.

\begin{table}
  \centering
  \caption{\label{Tb:STNmotif:PLexponent} Estimated exponents of the power-law behaviors in pre-event accumulation dynamics and post-event relaxation dynamics of the four financial variables: Volatility $v$, trading volume per minute $\omega_{\rm{cum}}$, average trade size per minute $\omega_{\rm{ave}}$, and turnover value per minute $f$.}
\medskip
\begin{tabular}{lccccc}
 \hline\hline
  $\kappa$ & $\beta_{\kappa,v}$ & $\beta_{\kappa,\omega_{\rm{cum}}}$ & $\beta_{\kappa,\omega_{\rm{ave}}}$ & $\beta_{\kappa,f}$ \\  %
 \hline
 \multicolumn{4}{l}{Panel A: Pre-event dynamics}& \smallskip
\\
  $L$  & $0.111(4)$ & $0.217(5)$ & $0.107(4)$ & $0.218(4)$ \\
  $M$  & $0.043(4)$ & $0.057(5)$ & $0.028(4)$ & $0.056(5)$ \\
  $S$  & $0.033(4)$ & $0.054(5)$ & $0.026(4)$ & $0.052(5)$\\
 \hline
 \multicolumn{4}{l}{Panel B: Post-event dynamics}& \smallskip
\\
  $L$  & $0.103(4)$ & $0.187(5)$ & $0.094(4)$ & $0.185(5)$ \\
  $M$  & $0.045(4)$ & $0.079(5)$ & $0.031(4)$ & $0.077(5)$ \\
  $S$  & $0.055(4)$ & $0.080(5)$ & $0.021(4)$ & $0.077(5)$ \\
 \hline\hline
 \end{tabular}
\end{table}

There are different possible explanations of the abnormal dynamics observed around the motif edges. One possible explanation is endogenous: The system evolves approaching a critical state through imitation and herding, which results in intermittent heavy trading activities. In such situation, the possibility to observe abnormal trading motifs increases. The exogenous explanation is that some collusive traders try to manipulate the price. We argue that both regimes coexist. Furthermore, it is intuitively rational to conjecture that large trades are more likely associated with manipulations than herding. A direct consequence is that the financial quantities accumulate faster than their relaxation such that $\beta_{L,x,{\rm{pre}}}>\beta_{L,x,{\rm{post}}}$. Similar reasoning can also explain $\beta_{L,x}>\beta_{M,x}>\beta_{S,x}$ for pre-event dynamics and $\beta_{L,x}>\beta_{M,x}\approx\beta_{S,x}$ for post-event dynamics. Certainly, detailed analysis is require to test these explanations and to explain $\beta_{\{M,S\},x,{\rm{pre}}}<\beta_{\{M,S\},x,{\rm{post}}}$, which is however beyond the scope of this Letter.

\section{Price impacts}

We investigate the price impact surrounding the transactions in the detected motifs though an event-study approach similar to the investigation of price impact of block trades \citep{Holthausen-Leftwich-Mayers-1990-JFE,Gemmill-1996-JF,Anderson-Cooper-Prevost-2006-FR,Frino-Jarnecic-Lepone-2007-ABACUS}. We calculate the raw trade-by-trade return series defined as the logarithmic difference of consecutive transaction prices. For each event $m\in{\mathcal{M}}$ where ${\mathcal{M}}$ is the set containing all transactions in the identified motifs, we extract a sequence $\{r_{m,i}:i=-10,\cdots,10\}$, where $r_{m,0}$ is the return between the transaction prices of the event and its preceding trade. For each event $m$, we designate a control set of 20 benchmark trades. The benchmark trades are selected from the trades of the same stock on other trading days. To avoid introducing other influence factors, each benchmark trade is also required to satisfy the following conditions: The benchmark trade has the same type ({\it{i.e.}} buyer-initialed or seller-initialed) as event $m$ and the transaction time of a benchmark trade must be close to that of the event when labeled in intraday time. For each benchmark trade $n$ in the control set, a sequence of trade-to-trade returns $\{r^{\rm{ben}}_{m,n,i}:i=-10,\cdots,10\}$ are determined similarly. The mean benchmark returns $r^{\rm{ben}}_{m,i}$ can be obtained as follows:
\begin{equation}
  r^{\rm{ben}}_{m,i} = \frac{1}{20}\sum_{n=1}^{20} r^{\rm{ben}}_{m,n,i}.
  \label{Eq:BenchmarkReturn}
\end{equation}
The excess return $r_t^{\rm{exc}}$ can be estimated by subtracting mean benchmark returns from the associated raw trade-to-trade returns,
\begin{equation}
  r^{\rm{exc}}_{m,i} = r_{m,i} - r^{\rm{ben}}_{m,i},~~i = -10, \cdots, 10.
  \label{Eq:ExcessReturn}
\end{equation}

\setlength\tabcolsep{2.5pt}
\begin{table*}[tb]
\centering%\small
 \caption{\label{Tb:EventReturn} Mean raw and excess returns 10 trades before and 10 trades after the buy-initiated and seller-initiated
 events. All the returns have been multiplied by a factor of $10^5$. Note that $^{*}$ indicates statistical significant at 5\% level, $^{**}$ indicates statistical significant at 1\% level, and $^{***}$ indicates statistical significant at 0.1\% level.}
 \medskip
 \centering
\begin{tabular}{r@{}lcr@{.}lcr@{.}lccr@{.}lcr@{.}lc}
 \hline\hline
   & && \multicolumn{6}{c}{Buyer-initiated events}&&\multicolumn{6}{c}{Seller-initiated events}\\  %
  \cline{4-9}  \cline{11-16}
 \multicolumn{2}{c}{$i$}& & \multicolumn{2}{c}{$\langle{r_{m,i}}\rangle$} & $p$-value &
 \multicolumn{2}{c}{$\langle{r^{\rm{exc}}_{m,i}}\rangle$} & $p$-value & & \multicolumn{2}{c}{$\langle{r_{m,i}}\rangle$} & $p$-value &
 \multicolumn{2}{c}{$\langle{r^{\rm{exc}}_{m,i}}\rangle$} & $p$-value \\
 \hline
 \multicolumn{8}{l}{Panel A: Trade-by-trade return}& \smallskip
\\
 $-$&10 && 3&84 & 0.493 & 4&38 & 0.437 && $-$14&92* & 0.028 & $-$12&71 & 0.073 \\
 $-$&9 && 6&80 & 0.220 & 9&39 & 0.097 && $-$2&09 & 0.740 & $-$0&60 & 0.927 \\
 $-$&8 && $-$0&19 & 0.972 & $-$0&18 & 0.974 && 3&35 & 0.596 & 4&12 & 0.524 \\
 $-$&7 && 8&62 & 0.136 & 12&18$^{*}$ & 0.040 && $-$3&93 & 0.495 & $-$2&02 & 0.732 \\
 $-$&6 && $-$3&69 & 0.570 & $-$1&85 & 0.778 && $-$0&99 & 0.874 & $-$2&37 & 0.715 \\
 $-$&5 && 18&99$^{**}$ & 0.001 & 20&87$^{***}$ & 0.001 && $-$1&94 & 0.763 & $-$1&29 & 0.846 \\
 $-$&4 && $-$7&65 & 0.186 & $-$6&48 & 0.276 && $-$2&12 & 0.726 & $-$4&11 & 0.517 \\
 $-$&3 && 4&75 & 0.428 & 6&10 & 0.324 && $-$14&49$^{*}$ & 0.023 & $-$15&62$^{*}$ & 0.017 \\
 $-$&2 && 11&77 & 0.083 & 17&61$^{*}$ & 0.010 && $-$2&91 & 0.609 & $-$6&70 & 0.262 \\
 $-$&1 && $-$6&49 & 0.324 & 5&50 & 0.411 && $-$9&72 & 0.157 & $-$22&45$^{**}$ & 0.001 \\
 &0 && 144&61$^{***}$ & 0.000 & 48&17$^{***}$ & 0.000 && $-$157&81$^{***}$ & 0.000 & $-$69&35$^{***}$ & 0.000 \\
 &1 && $-$78&37$^{***}$ & 0.000 & $-$2&69 & 0.740 && 95&05$^{***}$ & 0.000 & 27&96$^{**}$ & 0.001 \\
 &2 && 16&37$^{**}$ & 0.009 & 12&32 & 0.053 && $-$9&49 & 0.212 & $-$4&02 & 0.600 \\
 &3 && 1&85 & 0.766 & 1&27 & 0.843 && 5&66 & 0.417 & 8&73 & 0.219 \\
 &4 && 9&18 & 0.118 & 7&68 & 0.204 && $-$9&96 & 0.130 & $-$8&81 & 0.192 \\
 &5 && 5&69 & 0.333 & 8&59 & 0.155 && $-$8&00 & 0.273 & $-$7&52 & 0.311 \\
 &6 && $-$4&51 & 0.468 & $-$5&25 & 0.410 && -0&94 & 0.887 & $-$0&89 & 0.894 \\
 &7 && 9&89 & 0.111 & 11&40 & 0.076 && 1&55 & 0.800 & 1&73 & 0.783 \\
 &8 && 3&56 & 0.538 & 4&26 & 0.473 && $-$6&77 & 0.267 & $-$6&84 & 0.270 \\
 &9 && $-$0&77 & 0.895 & $-$1&33 & 0.823 && 11&90$^{*}$ & 0.049 & 9&31 & 0.140 \\
&10 && 5&83 & 0.292 & 5&04 & 0.376 && $-$9&87 & 0.087 & $-$8&99 & 0.129 \\\hline
 \multicolumn{8}{l}{Panel B: Multi-trade return}& \smallskip
\\
 $[-10$,&$-1]$ && 36&75$^{***}$ & 0.000 & 67&53$^{***}$ & 0.000 && $-$49&76$^{***}$ & 0.000 & $-$63&75$^{***}$ & 0.000 \\
 &0 && 144&61$^{***}$ & 0.000 & 48&17$^{***}$ & 0.000 && $-$157&81$^{***}$ & 0.000 & $-$69&35$^{***}$ & 0.000 \\
 $[1$,&$10]$ && -31&27$^{***}$ & 0.000 & 41&29$^{***}$ & 0.000 &&  69&12$^{***}$ & 0.000 & 10&66$^{***}$ & 0.000 \\
 \hline\hline
\end{tabular}
\end{table*}

Panel A of table \ref{Tb:EventReturn} presents the averages of the raw returns and the excess returns of buyer-initiated trades and seller-initiated trades, as well as the $p$-values of $t$-tests on the null hypothesis that an average is insignificantly different from 0. When $i=0$, the two average returns for buyer-initiated trades are positive and the two average returns for seller-initiated trades are negative. All these averages are significantly different from 0 at the significance level of 0.1\%. These values are comparable to those of all trades as shown in table 1 of \cite{Zhou-2012-QF}. There is a significant phenomenon of price reversal at $i=1$ after the events except for the excess returns after the buyer-initiated events, which is reminiscent of price reversal after large price movements \citep{Mu-Zhou-Chen-Kertesz-2010-NJP}. In addition, there are significantly positive average returns before buyer-initiated events and significantly negative average returns before seller-initiated events.

In Panel B of table \ref{Tb:EventReturn}, we present the cumulative sums of the trade-by-trade returns before and after the events. It is more evident that the raw cumulative return has the same sign as the return at $i=0$, which is followed by a significant price reversal. This finding seems normal and can be explained by endogenous herding behaviors. However, we observe a positive value (41.29) of the cumulative excess return after buyer-initiated events and a mild time reversal after seller-initiated events. These results can be explained in favor of possible price manipulations. When a group of collusive traders (or accounts) proceed to make profit by manipulating the price of a stock by the pump-and-dump strategy, they buy shares first and try to make their trades have larger price impact to pump the price. When dumping their shares, they would prefer to make their trades have ignorable price impact to hide their intention. This picture provide one explanation of why we observe continuous rise of the price after buyer-initiated trades and a mild time reversal after seller-initiated events.

\section{Conclusion}

In summary, we have studied three types of abnormal motifs in the trading networks of investors for 43 Chinese stocks. Abnormal dynamics of several financial variables (volatility, trading volume, average trade size, turnover value, and excess return) around the suspicious trades associated with the links of the identified motifs have been observed. The presence of price manipulation emerges as a possible explanation of these findings. Certainly, it does not rule out other possible explanations.

Our analysis provides a novel tools for the detection of trade-based price manipulation. It is not irrational to conjecture that there should be some traces left in the trading networks when price manipulations occur. Abnormal motifs are the basic building blocks of the interactions of stock pools. To provide conclusive evidence of manipulation, one need to perform further detailed analysis of the trading behaviors and the profits of the suspected traders. Our work paces the first step towards the uncovering of collusive manipulators. There is no doubt that techniques developed in this line will be helpful to market supervisors and policy makers.

%
%\begin{figure}
%  \begin{center}
%  \begin{minipage}{100mm}
%%    \subfigure[]{\resizebox*{5cm}{!}{\includegraphics{senu_gr1.eps}}\label{sample-figure_part_a}}%
%%    \subfigure[]{\resizebox*{5cm}{!}{\includegraphics{senu_gr2.eps}}\label{sample-figure_part_b}}%
%    \label{sample-figure}
%  \end{minipage}
%  \end{center}
%  \caption{Example of a two-part figure with individual sub-captions showing that all lines of figure captions range left. The parts should be referred to in the text as `figure~\ref{sample-figure}(a)' and `figure~\ref{sample-figure}(b)'.}
%\end{figure}
%
%
%\begin{table}
%\begin{center}
%\begin{minipage}{80mm}
%  \tbl{Radio-band beaming model parameters
%           for {FSRQs and BL Lacs.}}
%{\begin{tabular}{@{}lcccccc}\toprule
%   Class$^{\rm a}$
%  & $\gamma _1$ & $\gamma _2$$^{\rm b}$
%         & $\langle \gamma \rangle$
%         & $G$ & $f$ & $\theta _{c}$ \\
%\colrule
%   BL Lacs &5 & 36 & 7 & $-4.0$
%         & $1.0\times 10^{-2}$ & 10$^\circ$ \\
%   FSRQs & 5 & 40 & 11 & $-2.3$
%         & $0.5\times 10^{-2}$ & 14$^\circ$ \\
%   \botrule
%  \end{tabular}}
%\tabnote{$^{\rm a}$This is not as accurate, owing to numerical
%error.\\$^{\rm b}$An example table footnote to show the
%text turning over when a long footnote is
%inserted.}\label{symbols}
%\end{minipage}
%\end{center}
%\end{table}

\section*{Acknowledgements}

This work was partially supported by the National Natural Science Foundation of China (11075054, 71101052 and 71131007), Shanghai ``Chen Guang'' Project (2010CG32 and 2012CG34), Shanghai Rising Star (Follow-up) Program (11QH1400800), Program for Changjiang Scholars and Innovative Research Team in University (IRT1028), and the Fundamental Research Funds for the Central Universities.

\bibliographystyle{rQFL}
\bibliography{E:/Papers/Auxiliary/Bibliography}

\begin{thebibliography}{54}
\providecommand{\natexlab}[1]{#1}

\bibitem[\protect\citeauthoryear{Adamic
  {\itshape{et~al.}}}{2012}]{Adamic-Brunetti-Harris-Kirilenko-2012-SSRN}
Adamic, L., Brunetti, C., Harris, J. and Kirilenko, A., {Trading networks}
  [online]. , 2012. Available online at: http://ssrn.com/abstract=1361184
  (accessed 10 December 2012).

\bibitem[\protect\citeauthoryear{Allen and Gale}{1992}]{Allen-Gale-1992-RFS}
Allen, F. and Gale, D., {Stock-price manipulation}. {\itshape Rev. Financial
  Stud.}, 1992, \textbf{5}, 503--529.

\bibitem[\protect\citeauthoryear{Ammann and
  Kessler}{2009}]{Ammann-Kessler-2009-AFE}
Ammann, M. and Kessler, S.M., {Intraday characteristics of stock price
  crashes}. {\itshape Appl. Financial Econ.}, 2009, \textbf{19}, 1239--1255.

\bibitem[\protect\citeauthoryear{Anderson
  {\itshape{et~al.}}}{2006}]{Anderson-Cooper-Prevost-2006-FR}
Anderson, H.D., Cooper, S. and Prevost, A.K., {Block trade price asymmetry and
  changes in depth: Evidence from the Australian Stock Exchange}. {\itshape
  Financial Rev.}, 2006, \textbf{41}, 247--271.

\bibitem[\protect\citeauthoryear{Baker}{1981}]{Baker-1981-PhD}
Baker, W.E., {\itshape Markets as Networks: A Multimethod Study of Trading
  Networks in a Securities Market} 1981  (Northwestern University: Evanston),
  PhD Thesis.

\bibitem[\protect\citeauthoryear{Baker}{1984}]{Baker-1984-AJS}
Baker, W.E., {The social-structure of a national securities market}. {\itshape
  Amer. J. Soc.}, 1984, \textbf{89}, 775--811.

\bibitem[\protect\citeauthoryear{Clauset
  {\itshape{et~al.}}}{2009}]{Clauset-Shalizi-Newman-2009-SIAMR}
Clauset, A., Shalizi, C.R. and Newman, M.E.J., {Power-law distributions in
  empirical data}. {\itshape SIAM Rev.}, 2009, \textbf{51}, 661--703.

\bibitem[\protect\citeauthoryear{Comerton-Forde and
  Putni{\c{n}}{\v{s}}}{2011}]{ComertonForde-Putnins-2011-JFI}
Comerton-Forde, C. and Putni{\c{n}}{\v{s}}, T.J., {Measuring closing price
  manipulation}. {\itshape J. Finan. Intermediation}, 2011, \textbf{20},
  135--158.

\bibitem[\protect\citeauthoryear{Cumming and
  Johan}{2008}]{Cumming-Johan-2008-ALER}
Cumming, D. and Johan, S., {Global market surveillance}. {\itshape Amer. Law
  Econ. Rev.}, 2008, \textbf{10}, 454--506.

\bibitem[\protect\citeauthoryear{Diaz
  {\itshape{et~al.}}}{2011}]{Diaz-Theodoulidis-Sampaio-2011-ESA}
Diaz, D., Theodoulidis, B. and Sampaio, P., {Analysis of stock market
  manipulations using knowledge discovery techniques applied to intraday trade
  prices}. {\itshape Expert Sys. Appl.}, 2011, \textbf{38}, 12757--12771.

\bibitem[\protect\citeauthoryear{Erenburg and
  Lasser}{2009}]{Erenburg-Lasser-2009-RFE}
Erenburg, G. and Lasser, D., {Electronic limit order book and order submission
  choice around macroeconomic news}. {\itshape Rev. Financial Econ.}, 2009,
  \textbf{18}, 172--182.

\bibitem[\protect\citeauthoryear{Felixson and
  Pelli}{1999}]{Felixson-Pelli-1999-JMFM}
Felixson, K. and Pelli, A., {Day end returns--stock price manipulation}.
  {\itshape J. Multinat. Financial Manag.}, 1999, \textbf{9}, 95--127.

\bibitem[\protect\citeauthoryear{Fleming and
  Remolona}{1997}]{Fleming-Remolona-1997-EPR}
Fleming, M.J. and Remolona, E.M., {What moves the bond market?}. {\itshape
  Econ. Policy Rev.}, 1997, pp. 31--50.

\bibitem[\protect\citeauthoryear{Fleming and
  Remolona}{1999}]{Fleming-Remolona-1999-JF}
Fleming, M.J. and Remolona, E.M., {Price formation and liquidity in the U.S.
  treasury market: the response to public information}. {\itshape J. Finance},
  1999, \textbf{54}, 1901--1915.

\bibitem[\protect\citeauthoryear{Franke
  {\itshape{et~al.}}}{2006}]{Franke-GeyerSchulz-Hoser-2006-bk}
Franke, M., Geyer-Schulz, A. and Hoser, B., On the analysis of asymmetric
  directed communication structures in electronic election markets. In
  {\itshape Agent-Based Computational Modelling}, edited by F.C. Billari,
  T.~Fent, A.~Prskawetz and J.~Scheffran, pp. 37--59, 2006  (Physica-Verlag HD:
  Heidelberg).

\bibitem[\protect\citeauthoryear{Franke
  {\itshape{et~al.}}}{2008}]{Franke-Hoser-Schroder-2008-bk}
Franke, M., Hoser, B. and Schr{\"{o}}der, J., On the analysis of irregular
  stock market trading behavior. In {\itshape Data Analysis, Machine Learning
  and Applications}, edited by C.~Preisach, H.~Burkhardt, L.~Schmidt-Thieme and
  R.~Decker, pp. 355--362, 2008  (Springer: Berlin).

\bibitem[\protect\citeauthoryear{Frino
  {\itshape{et~al.}}}{2007}]{Frino-Jarnecic-Lepone-2007-ABACUS}
Frino, A., Jarnecic, E. and Lepone, A., {The determinants of the price impact
  of block trades: Further evidence}. {\itshape ABACUS}, 2007, \textbf{43},
  94--106.

\bibitem[\protect\citeauthoryear{Gemmill}{1996}]{Gemmill-1996-JF}
Gemmill, G., {Transparency and liquidity: A study of block trades in the London
  Stock Exchange under different publication rules}. {\itshape J. Finance},
  1996, \textbf{51}, 1765--1790.

\bibitem[\protect\citeauthoryear{Gowda and
  Krishna}{1978}]{Gowda-Krishna-1978-PtnR}
Gowda, K.C. and Krishna, G., {Agglomerative clustering using the concept of
  mutual nearest neighbourhood}. {\itshape Pattern Recognition}, 1978,
  \textbf{10}, 105--112.

\bibitem[\protect\citeauthoryear{Gro{\ss}-Klu{\ss}mann and
  Hautsch}{2009}]{GrossKlussmann-Hautsch-2009-SSRN}
Gro{\ss}-Klu{\ss}mann, A. and Hautsch, N., {Quantifying high-frequency market
  reactions to real-time news sentiment announcements}  [online]. , 2009.
  Available online at: http://ssrn.com/abstract=1536005 (accessed Oct. 3,
  2012).

\bibitem[\protect\citeauthoryear{Hamelink}{2003}]{Hamelink-2003-JFc}
Hamelink, F., {Systematic patterns before and after large price changes:
  evidence from high frequency data from the Paris Bourse}. {\itshape J.
  Forecast.}, 2003, \textbf{22}, 533--549.

\bibitem[\protect\citeauthoryear{Hillion and
  Suominen}{2004}]{Hillion-Suominen-2004-JFinM}
Hillion, P. and Suominen, M., {The manipulation of closing prices}. {\itshape
  J. Financial Markets}, 2004, \textbf{7}, 351--375.

\bibitem[\protect\citeauthoryear{Holthausen
  {\itshape{et~al.}}}{1990}]{Holthausen-Leftwich-Mayers-1990-JFE}
Holthausen, R.W., Leftwich, R.W. and Mayers, D., {Large-block transactions, the
  speed of response, and temporary and permanent stock-price effects}.
  {\itshape J. Financial Econ.}, 1990, \textbf{26}, 71--95.

\bibitem[\protect\citeauthoryear{Islam
  {\itshape{et~al.}}}{2009}]{Islam-Haque-Alam-Tarikuzzaman-2009-ICCIT}
Islam, M.N., Haque, S.M.R., Alam, K.M. and Tarikuzzaman, M., {An approach to
  improve collusion set detection using MCL algorithm}. In {\itshape
  Proceedings of the }{\itshape Proceedings of the 12th International
  Conference on Computer and Information Technology}, pp. 237--242, 2009
  (ICCIT 2009 Conference Secretariat: Dhaka).

\bibitem[\protect\citeauthoryear{Jarvis and
  Patrick}{1973}]{Jarvis-Patrick-1973-IEEEtc}
Jarvis, R.A. and Patrick, E.A., {Clustering using a similarity measure based on
  shared near neighbors}. {\itshape IEEE Trans. Comput.}, 1973, \textbf{22},
  1025--1034.

\bibitem[\protect\citeauthoryear{Jiang
  {\itshape{et~al.}}}{2005}]{Jiang-Mahoney-Mei-2005-JFE}
Jiang, G.L., Mahoney, P.G. and Mei, J.P., {Market manipulation: A comprehensive
  study of stock pools}. {\itshape J. Financial Econ.}, 2005, \textbf{77},
  147--170.

\bibitem[\protect\citeauthoryear{Jiang
  {\itshape{et~al.}}}{2009}]{Jiang-Chen-Zhou-2009-PA}
Jiang, Z.Q., Chen, W. and Zhou, W.X., {Detrended fluctuation analysis of
  intertrade durations}. {\itshape Physica A}, 2009, \textbf{388}, 433--440.

\bibitem[\protect\citeauthoryear{Jiang and Zhou}{2010}]{Jiang-Zhou-2010-PA}
Jiang, Z.Q. and Zhou, W.X., {Complex stock trading network among investors}.
  {\itshape Physica A}, 2010, \textbf{389}, 4929--4941.

\bibitem[\protect\citeauthoryear{Joulin
  {\itshape{et~al.}}}{2008}]{Joulin-Lefevre-Grunberg-Bouchaud-2008-Wilmott}
Joulin, A., Lefevre, A., Grunberg, D. and Bouchaud, J.P., {Stock price jumps:
  News and volume play a minor role}. {\itshape Wilmott Magazine}, 2008,
  \textbf{Sep/Oct}, 1--7.

\bibitem[\protect\citeauthoryear{Kovanen
  {\itshape{et~al.}}}{2011}]{Kovanen-Karsai-Kaski-Kertesz-Saramaki-2011-JSM}
Kovanen, L., Karsai, M., Kaski, K., Kert{\'{e}}sz, J. and Saram{\"{a}}ki, J.,
  {Temporal motifs in time-dependent networks}. {\itshape J. Stat. Mech.},
  2011, \textbf{2011}, P11005.

\bibitem[\protect\citeauthoryear{Lillo
  {\itshape{et~al.}}}{2008}]{Lillo-Moro-Vaglica-Mantegna-2008-NJP}
Lillo, F., Moro, E., Vaglica, G. and Mantegna, R.N., {Specialization and
  herding behavior of trading firms in a financial market}. {\itshape New J.
  Phys.}, 2008, \textbf{10}, 043019.

\bibitem[\protect\citeauthoryear{Mahoney}{1999}]{Mahoney-1999-JFE}
Mahoney, P.G., {The stock pools and the Securities Exchange Act}. {\itshape J.
  Financial Econ.}, 1999, \textbf{51}, 343--369.

\bibitem[\protect\citeauthoryear{Milo
  {\itshape{et~al.}}}{2004}]{Milo-Itzkovitz-Kashtan-Levitt-ShenOrr-Ayzenshtat-Sheffer-Alon-2004-Science}
Milo, R., Itzkovitz, S., Kashtan, N., Levitt, R., Shen-Orr, S., Ayzenshtat, I.,
  Sheffer, M. and Alon, U., {Superfamilies of evolved and designed networks}.
  {\itshape Science}, 2004, \textbf{303}, 1538--1542.

\bibitem[\protect\citeauthoryear{Milo
  {\itshape{et~al.}}}{2002}]{Milo-ShenOrr-Itzkovitz-Kashtan-Chklovskii-Alon-2002-Science}
Milo, R., Shen-Orr, S., Itzkovitz, S., Kashtan, N., Chklovskii, D. and Alon,
  U., {Network motifs: Simple building blocks of complex networks}. {\itshape
  Science}, 2002, \textbf{298}, 824--827.

\bibitem[\protect\citeauthoryear{Mu
  {\itshape{et~al.}}}{2010}]{Mu-Zhou-Chen-Kertesz-2010-NJP}
Mu, G.H., Zhou, W.X., Chen, W. and Kert{\'e}sz, J., {Order flow dynamics around
  extreme price changes on an emerging stock market}. {\itshape New J. Phys.},
  2010, \textbf{12}, 075037.

\bibitem[\protect\citeauthoryear{{\"{O}}{\u{g}}{\"{u}}t
  {\itshape{et~al.}}}{2009}]{Ogut-Doganay-Aktas-2009-ESA}
{\"{O}}{\u{g}}{\"{u}}t, H., Do{\u{g}}anay, M.M. and Akta{\c{s}}, R., {Detecting
  stock-price manipulation in an emerging market: The case of Turkey}.
  {\itshape Expert Sys. Appl.}, 2009, \textbf{36}, 11944--11949.

\bibitem[\protect\citeauthoryear{Palshikar and
  Apte}{2008}]{Palshikar-Apte-2008-DMKD}
Palshikar, G.K. and Apte, M.M., {Collusion set detection using graph
  clustering}. {\itshape Data Min. Knowl. Disc.}, 2008, \textbf{16}, 135--164.

\bibitem[\protect\citeauthoryear{Ponzi
  {\itshape{et~al.}}}{2009}]{Ponzi-Lillo-Mantegna-2009-PRE}
Ponzi, A., Lillo, F. and Mantegna, R.N., {Market reaction to a bid-ask spread
  change: A power-law relaxation dynamics}. {\itshape Phys. Rev. E}, 2009,
  \textbf{80}, 016112.

\bibitem[\protect\citeauthoryear{Putni{\c{n}}{\v{s}}}{2012}]{Putnins-2012-JES}
Putni{\c{n}}{\v{s}}, T.J., {Market manipulation: A survey}. {\itshape J. Econ.
  Surveys}, 2012, \textbf{26}, 952--967.

\bibitem[\protect\citeauthoryear{Sun
  {\itshape{et~al.}}}{2010}]{Sun-Cheng-Shen-Wang-2010-PP}
Sun, X.Q., Cheng, X.Q., Shen, H.W. and Wang, Z.Y., {Statistical properties of
  trading activity in Chinese Stock Market}. {\itshape Phys. Proc.}, 2010,
  \textbf{3}, 1699--1706.

\bibitem[\protect\citeauthoryear{Sun
  {\itshape{et~al.}}}{2011}]{Sun-Cheng-Shen-Wang-2011-PA}
Sun, X.Q., Cheng, X.Q., Shen, H.W. and Wang, Z.Y., {Distinguishing manipulated
  stocks via trading network analysis}. {\itshape Physica A}, 2011,
  \textbf{390}, 3427--3434.

\bibitem[\protect\citeauthoryear{Sun
  {\itshape{et~al.}}}{2012}]{Sun-Shen-Cheng-Wang-2012-PLoS1}
Sun, X.Q., Shen, H.W., Cheng, X.Q. and Wang, Z.Y., {Degree-strength correlation
  reveals anomalous trading behavior}. {\itshape PLoS One}, 2012, \textbf{7},
  e45598.

\bibitem[\protect\citeauthoryear{T{\'o}th
  {\itshape{et~al.}}}{2009}]{Toth-Kertesz-Farmer-2009-EPJB}
T{\'o}th, B., Kert{\'e}sz, J. and Farmer, J.D., {Studies of the limit order
  book around large price changes}. {\itshape Eur. Phys. J. B}, 2009,
  \textbf{71}, 499--510.

\bibitem[\protect\citeauthoryear{Tseng
  {\itshape{et~al.}}}{2009}]{Tseng-Li-Chen-Wang-2009-ACS}
Tseng, J.J., Li, S.P., Chen, S.H. and Wang, S.C., {Emergence of scale-free
  networks in markets}. {\itshape Adv. Complex Sys.}, 2009, \textbf{12},
  87--97.

\bibitem[\protect\citeauthoryear{Tseng
  {\itshape{et~al.}}}{2010{\natexlab{a}}}]{Tseng-Li-Wang-2010-EPJB}
Tseng, J.J., Li, S.P. and Wang, S.C., {Experimental evidence for the interplay
  between individual wealth and transaction network}. {\itshape Eur. Phys. J.
  B}, 2010{\natexlab{a}}, \textbf{73}, 69--74.

\bibitem[\protect\citeauthoryear{Tseng
  {\itshape{et~al.}}}{2010{\natexlab{b}}}]{Tseng-Lin-Lin-Wang-Li-2010-PA}
Tseng, J.J., Lin, C.H., Lin, C.T., Wang, S.C. and Li, S.P., {Statistical
  properties of agent-based models in markets with continuous double auction
  mechanism}. {\itshape Physica A}, 2010{\natexlab{b}}, \textbf{389},
  1699--1707.

\bibitem[\protect\citeauthoryear{Tumminello
  {\itshape{et~al.}}}{2012}]{Tumminello-Lillo-Piilo-Mantegna-2012-NJP}
Tumminello, M., Lillo, F., Piilo, J. and Mantegna, R.N., {Identification of
  clusters of investors from their real trading activity in a financial
  market}. {\itshape New J. Phys.}, 2012, \textbf{14}, 013041.

\bibitem[\protect\citeauthoryear{Wang
  {\itshape{et~al.}}}{2011}]{Wang-Zhou-Guan-2011-PA}
Wang, J.J., Zhou, S.G. and Guan, J.H., {Characteristics of real futures trading
  networks}. {\itshape Physica A}, 2011, \textbf{390}, 398--409.

\bibitem[\protect\citeauthoryear{Wang
  {\itshape{et~al.}}}{2012}]{Wang-Zhou-Guan-2012-Nc}
Wang, J.J., Zhou, S.G. and Guan, J.H., {Detecting potential collusive cliques
  in futures markets based on trading behaviors from real data}. {\itshape
  Neurocomputing}, 2012, \textbf{92}, 44--53.

\bibitem[\protect\citeauthoryear{Wang
  {\itshape{et~al.}}}{2008}]{Wang-Tseng-Tai-Lai-Wu-Chen-Li-2008-EPJB}
Wang, S.C., Tseng, J.J., Tai, C.C., Lai, K.H., Wu, W.S., Chen, S.H. and Li,
  S.P., {Network topology of an experimental futures exchange}. {\itshape Eur.
  Phys. J. B}, 2008, \textbf{62}, 105--111.

\bibitem[\protect\citeauthoryear{Zawadowski
  {\itshape{et~al.}}}{2006}]{Zawadowski-Andor-Kertesz-2006-QF}
Zawadowski, A.G., Andor, G. and Kert{\'e}sz, J., {Short-term market reaction
  after extreme price changes of liquid stocks}. {\itshape Quant. Finance},
  2006, \textbf{6}, 283--295.

\bibitem[\protect\citeauthoryear{Zawadowski
  {\itshape{et~al.}}}{2004}]{Zawadowski-Kertesz-Andor-2004-PA}
Zawadowski, A.G., Kert{\'e}sz, J. and Andor, G., {Large price changes on small
  scales}. {\itshape Physica A}, 2004, \textbf{344}, 221--226.

\bibitem[\protect\citeauthoryear{Zhou}{2012}]{Zhou-2012-QF}
Zhou, W.X., {Universal price impact functions of individual trades in an
  order-driven market}. {\itshape Quant. Finance}, 2012, \textbf{12},
  1253--1263.

\bibitem[\protect\citeauthoryear{Zhou
  {\itshape{et~al.}}}{2012}]{Zhou-Mu-Kertesz-2012-NJP}
Zhou, W.X., Mu, G.H. and Kert{\'e}sz, J., {Random matrix approach to the
  dynamics of stock inventory variations}. {\itshape New J. Phys.}, 2012,
  \textbf{14}, 093025.

\end{thebibliography}
%\bibliography{/home/zqjiang/research/Papers/Auxiliary/Bibliography}

\end{document}